\documentclass[prd,aps,floats,twocolumn,showpacs,superscriptaddress,preprintnumbers,showkeys]{revtex4-1}
\usepackage{multirow,amssymb,amsmath,feynmf,subfigure,pdfpages,placeins,mathrsfs,slashed}
\usepackage{graphicx}
\usepackage{dcolumn}

\newcolumntype{d}[1]{D{.}{.}{#1}}
\newcolumntype{F}[1]{D{@}{}{#1}}

\newcommand{\overbar}[1]{\mkern 1.5mu\overline{\mkern-1.5mu#1\mkern-1.5mu}\mkern 1.5mu}

\graphicspath{{/FeynmanDiagrams/}}

\begin{document}

\preprint{
\vbox{
\hbox{ADP-15-27/T929}
\hbox{Edinburgh 2015/17}
\hbox{DESY 15-150}
}}

\title[SU(3) breaking in hyperon transition vector form factors]{SU(3) breaking in hyperon transition vector form factors}
\author{P.E.~Shanahan}\affiliation{ARC Centre of Excellence in Particle Physics at the Terascale and CSSM, Department of Physics, University of Adelaide, Adelaide SA 5005, Australia}
\author{A.N.~Cooke}\affiliation{School of Physics and Astronomy, University of Edinburgh, Edinburgh EH9 3JZ, UK}
\author{R.~Horsley}\affiliation{School of Physics and Astronomy, University of Edinburgh, Edinburgh EH9 3JZ, UK}
\author{Y.~Nakamura}\affiliation{RIKEN Advanced Institute for Computational Science, Kobe, Hyogo 650-0047, Japan}
\author{P.E.L.~Rakow}\affiliation{Theoretical Physics Division, Department of Mathematical Sciences, University of Liverpool, Liverpool L69 3BX, UK}
\author{G.~Schierholz}\affiliation{Deutsches Elektronen-Synchrotron DESY, 22603 Hamburg, Germany}
\author{A.W.~Thomas}\affiliation{ARC Centre of Excellence in Particle Physics at the Terascale and CSSM, Department of Physics, University of Adelaide, Adelaide SA 5005, Australia}
\author{R.D.~Young}\affiliation{ARC Centre of Excellence in Particle Physics at the Terascale and CSSM, Department of Physics, University of Adelaide, Adelaide SA 5005, Australia}
\author{J.M.~Zanotti}\affiliation{ARC Centre of Excellence in Particle Physics at the Terascale and CSSM, Department of Physics, University of Adelaide, Adelaide SA 5005, Australia}

\begin{abstract}
We present a calculation of the SU(3)-breaking corrections to the hyperon transition vector form factors to $\mathcal{O}(p^4)$ in heavy baryon chiral perturbation theory with finite-range regularisation. Both octet and decuplet degrees of freedom are included. 
We formulate a chiral expansion at the kinematic point $Q^2=-(M_{B_1}-M_{B_2})^2$, which can be conveniently accessed in lattice QCD. The two unknown low-energy constants at this point are constrained by lattice QCD simulation results for the $\Sigma^-\rightarrow n$ and $\Xi^0\rightarrow \Sigma^+$ transition form factors. Hence we determine lattice-informed values of $f_1$ at the physical point. This work constitutes progress towards the precise determination of $|V_{us}|$ from hyperon semileptonic decays.
\end{abstract}



\maketitle

\section{Introduction}

The Cabibbo-Kobayashi-Maskawa (CKM) quark-mixing matrix elements are fundamental parameters of the Standard Model. The precise determination of these quantities, which encode the flavor structure of the quark sector, is thus of great importance~\cite{Antonelli:2009ws}. In particular, a stringent test of CKM unitarity~\cite{Cirigliano:2009wk} is given by the first-row relation $|V_{ud}|^2+|V_{us}|^2+|V_{ub}|^2=1$. The matrix element $|V_{us}|$ has traditionally been extracted from measurements of kaon semileptonic and leptonic decays and the hadronic decays of tau leptons. These determinations have (until recently~\cite{Boyle:2013xya,MaltmanPrep2,Maltmanprep}) been in slight tension~\cite{pdg}. A further independent extraction of $|V_{us}|$ can be performed based on hyperon beta decay. In particular, the product $|V_{us}f_1(Q^2=0)|$ can be extracted from experiment at the percent-level~\cite{Cabibbo:2003cu}; precise calculations of the hadronic corrections to the vector form factors $f_1(Q^2=0)$ are therefore of great interest.

While the Ademollo-Gatto theorem~\cite{PhysRevLett.13.264} protects the vector form factors from leading SU(3)-symmetry--breaking corrections generated by the mass difference of the strange and nonstrange quarks, knowledge of the second-order breaking corrections to $f_1(Q^2=0)$ is crucial to obtain a precise value of $|V_{us}|$~\cite{Cabibbo:2003cu,Mateu:2005wi}. Estimates of these corrections have been performed based on quark models~\cite{Donoghue:1986th,Schlumpf:1994fb}, $1/N_c$ expansions~\cite{PhysRevD.58.094028}, chiral effective field theory~\cite{Villadoro:2006nj,Lacour:2007wm,Geng:2014efa,Anderson:1993as,Kaiser:2001yc} and quenched and unquenched lattice QCD~\cite{Guadagnoli200763,PhysRevD.79.074508,PhysRevD.86.114502}. An enduring puzzle has been that the sign of the SU(3) breaking corrections determined in lattice QCD (at this stage away from the physical pseudoscalar masses) and quark models is, in general, opposite to that determined from relativistic and heavy baryon chiral perturbation theory and $1/N_c$ expansions.

Using finite-range regularised chiral perturbation theory, which appears to offer markedly improved convergence properties of the (traditionally poorly convergent~\cite{Donoghue:1998bs,Borasoy:2002jv}) SU(3) chiral expansion, we revisit the effective field theory estimates of the SU(3) breaking in $f_1$.
To enable us to use 2+1-flavor lattice QCD to inform this determination, we choose to investigate the form factors at $Q^2=-(M_{B_1}-M_{B_2})^2$ (i.e., $\vec{q}=0$ in our lattice simulations with fixed zero sink momentum) instead of at $Q^2=0$ as is standard. This choice avoids the need to model the $Q^2$-dependence of the lattice simulation results for $f_1$ and correct to $Q^2=0$ before chiral and finite-volume corrections have been applied. This small shift in $Q^2$ would be highly model-dependent with our simulation results; for example, after using linear and dipole interpolations to $Q^2=0$ the results differ by more than the statistical uncertainty.  

After extrapolation to the physical regime we use similar parameterisations to those used to fit experimental results to extrapolate to $Q^2=0$, allowing comparison with other works. 
The sign of the SU(3)-symmetry-breaking corrections that we find is consistent with the results of quark models.
Furthermore, while we have lattice simulation results for only the $\Sigma^-\rightarrow n$ and $\Xi^0\rightarrow\Sigma^+$ transitions, our chiral extrapolations also allow us to extract (lattice-informed) results for the SU(3)-breaking in the $\Lambda\rightarrow p$ and $\Xi^- \rightarrow \Lambda$ transitions.

\section{Chiral perturbation theory}
\label{sec:ChiPT}

We use a formalism based on heavy-baryon chiral perturbation theory, with finite-range regularisation, to perform chiral and infinite-volume extrapolations of lattice simulation results for $f_1$.
Our chiral expansion is different from those used previously~\cite{Villadoro:2006nj,Lacour:2007wm,Geng:2014efa,Anderson:1993as,Kaiser:2001yc} in that we consider the form factor at $Q^2_\text{max}=-(M_{B_1}-M_{B_2})^2$ (corresponding to $\vec{q}=0$ in our lattice simulations), rather than $Q^2=0$. In addition to removing the model dependence of a $Q^2\rightarrow 0$ interpolation, a feature of this approach is that several free low-energy constants appear in the chiral expansion (while $f_1(Q^2=0)$ is protected from such coefficients to this order by Ademollo-Gatto~\cite{PhysRevLett.13.264}). These free constants are fit to the lattice QCD simulation results. In practice, fitting these terms to the lattice simulation results may also absorb the effect of potentially significant higher-order terms neglected in the conventional chiral expansion.

The structure of the master formula for our chiral extrapolation to $\mathcal{O}(p^4)$ is
\begin{align}\nonumber
f_1^{B_1B_2}(Q^2=(M_B&-M_{B'})^2) \\\nonumber
= & f_1^\text{SU(3)} + C_{B_1B_2}Q^2 + \text{(Tadpole Loops)} \\ \nonumber
& + \text{(Octet Loops)} + \text{(Decuplet Loops)}\\\label{eq:Master}
& + (1/M_0 \text{ corrections}).
\end{align}
The terms $f_1^\text{SU(3)}$ are the SU(3)-symmetric values of the vector form factors. 
At this order, the only undetermined chiral coefficients (which we will fit to the lattice simulation results) appear in the term proportional to $Q^2$. These are the same coefficients which appear in expressions for the octet baryon electric radii (labelled $b_D$ and $b_F$ in Ref.~\cite{Shanahan:2014cga}, for example). We discuss each remaining term of Eq.~\eqref{eq:Master} in turn.

\subsection{Terms linear in $Q^2$}

The term in Eq.~\eqref{eq:Master} which is linear in $Q^2$, (i.e., for the application to our lattice simulations with $\vec{q}=0$ it is proportional to $(M_{B_1}-M_{B_2})^2$), comes from the Lagrangian piece
\begin{equation}
\mathcal{L}=-ev^\mu(D^\nu F_{\mu\nu}^+)\left[ 2b_D \text{Tr} \overline{B} \left\{\lambda, B \right\} + 2b_F \text{Tr} \overline{B} \left[\lambda, B \right]\right],
\end{equation}
which is equivalent to Eq.~(4) in Ref.~\cite{Shanahan:2014cga} (where we have dropped the term proportional to $\text{Tr}\lambda$ which vanishes here). Here $\lambda$ is the strangeness-changing matrix
\begin{equation} \left(
\begin{matrix}
0 & 0 & 1 \\
0 & 0 & 0 \\
0 & 0 & 0 
\end{matrix} \right).
\end{equation}
The two undetermined coefficients $b_D$ and $b_F$, which appear in the coefficients $C_{B_1B_2}$ which are tabulated in appendix~\ref{sec:CoeffTabs}, 
will be fit to the lattice simulation results.
We note that in Ref.~\cite{Shanahan:2014cga}, the identically-named parameters are interpreted as chiral limit form factors at fixed values of $Q^2$ (that is, they may in principle vary with $Q^2$). 
In the present work, these terms may be interpreted as genuine chiral low-energy constants, i.e., the $Q^2=0$ limit of those in Ref.~\cite{Shanahan:2014cga}.

\subsection{Loop diagram contributions}

\begin{figure*}
\centering
\subfigure[]{\label{fig:RenOct}
\includegraphics[width=0.2\textwidth]{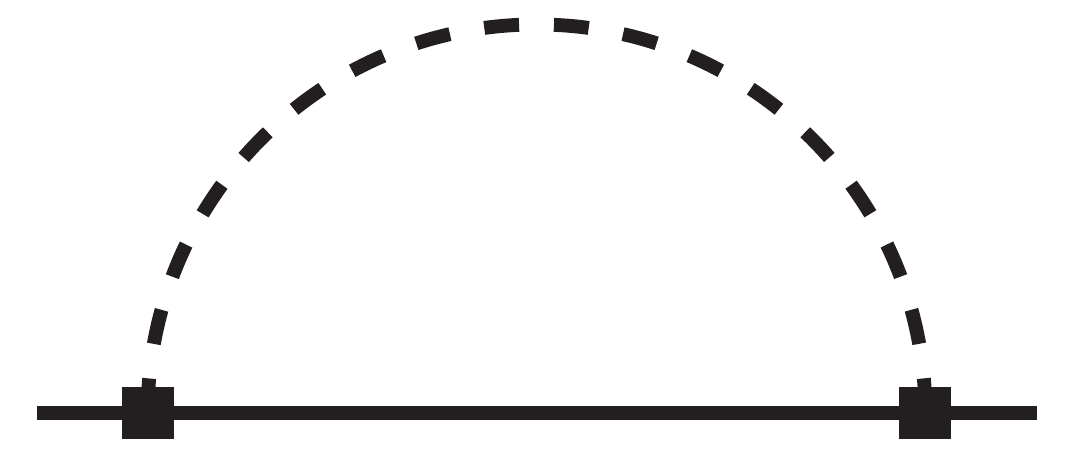}}
\subfigure[]{\label{fig:RenDec}
\includegraphics[width=0.2\textwidth]{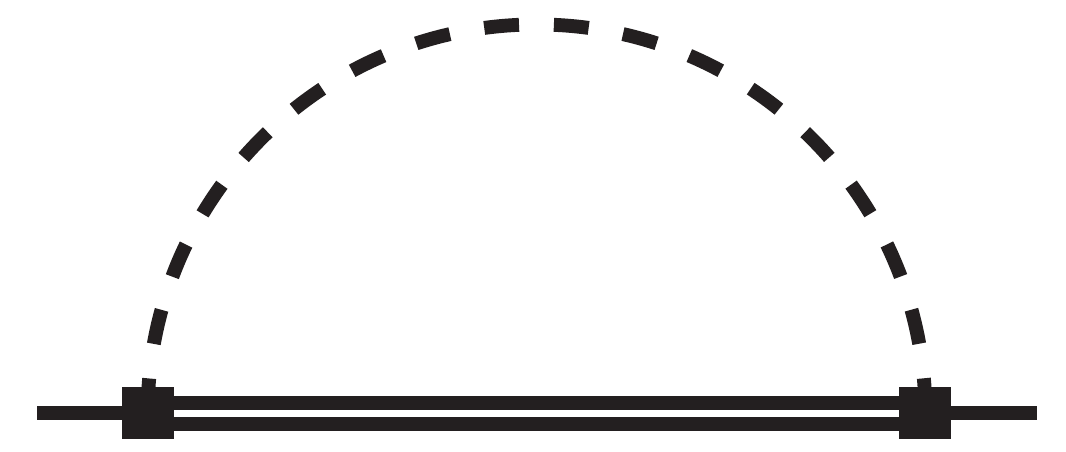}}
\subfigure[]{\label{fig:OctCrossLoop}
\includegraphics[width=0.2\textwidth]{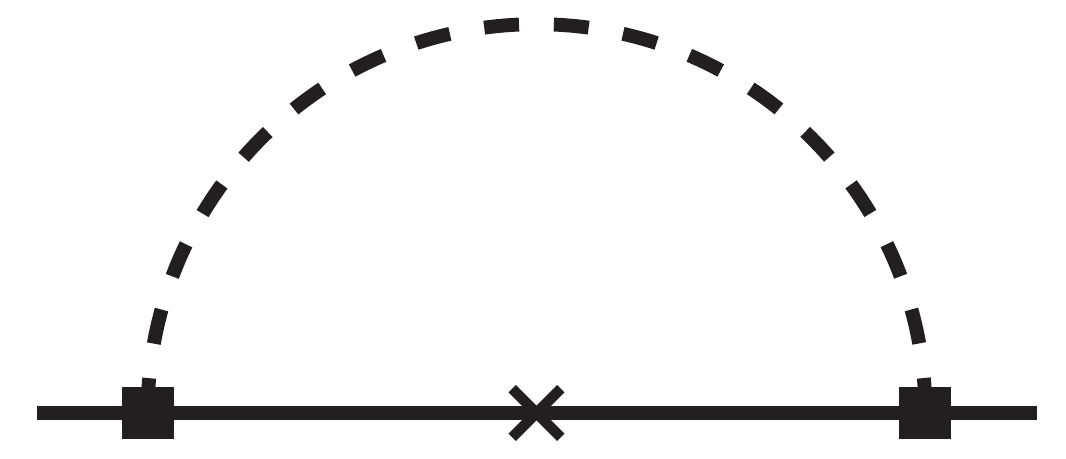}}
\subfigure[]{\label{fig:DecCrossLoop}
\includegraphics[width=0.2\textwidth]{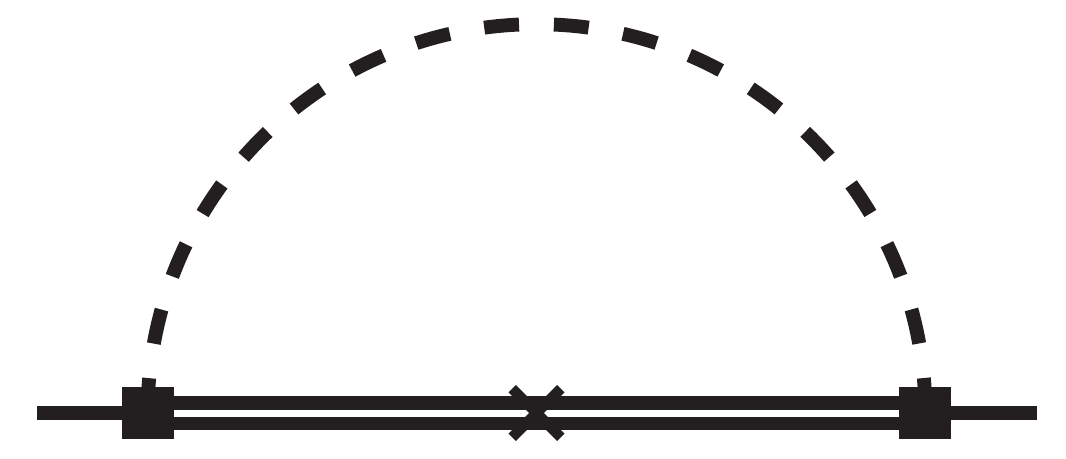}}
\subfigure[]{\label{fig:MesInsCrossLoopOct}
\includegraphics[width=0.2\textwidth]{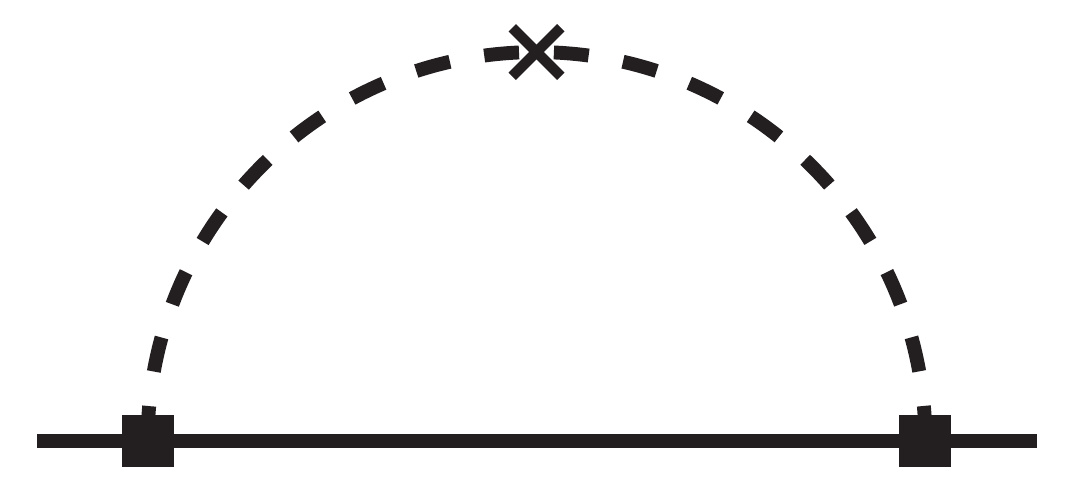}}
\subfigure[]{\label{fig:MesInsCrossLoopDec}
\includegraphics[width=0.2\textwidth]{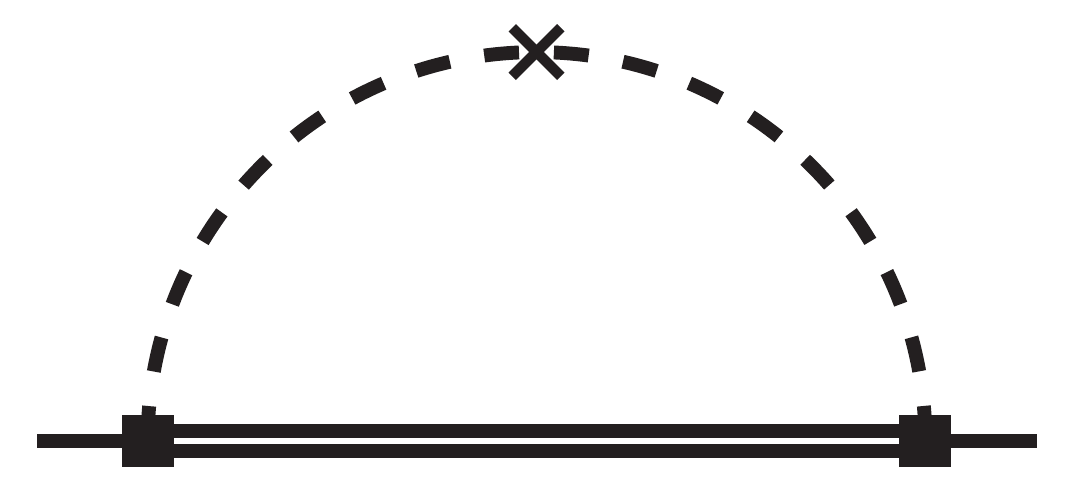}}
\subfigure[]{\label{fig:CrossTadLoop}
\includegraphics[width=0.14\textwidth]{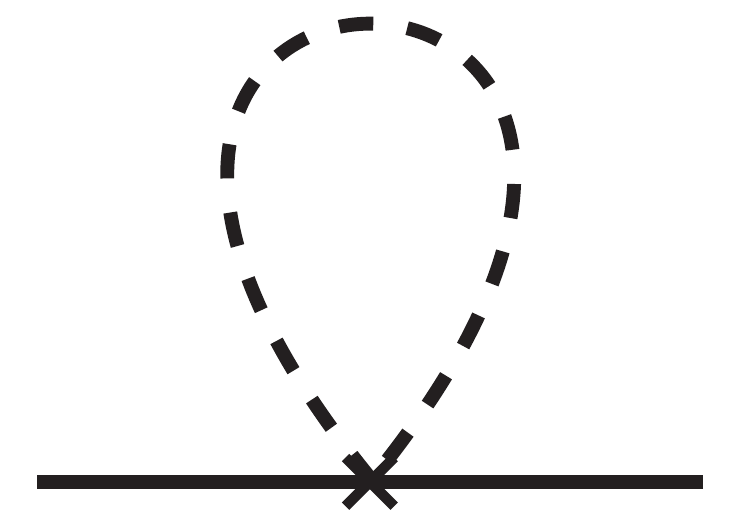}}
\subfigure[]{\label{fig:MesInsCrossTadLoop}
\includegraphics[width=0.14\textwidth]{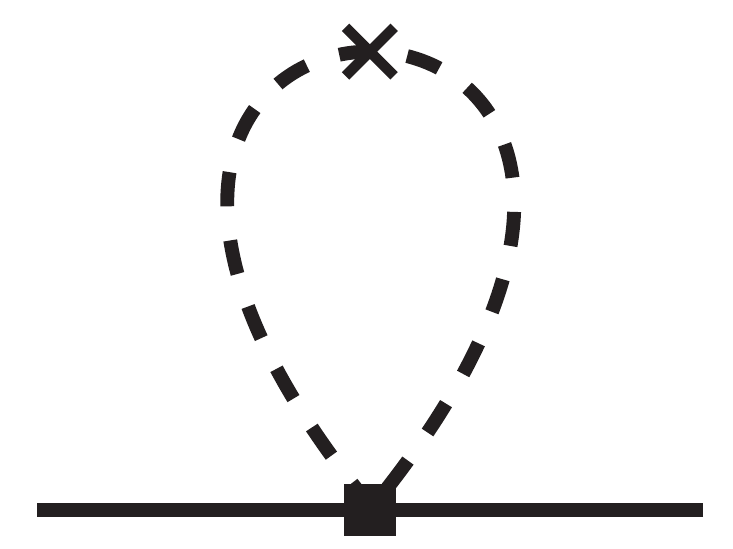}}
\subfigure[]{\label{fig:RenVertC}
\includegraphics[width=0.2\textwidth]{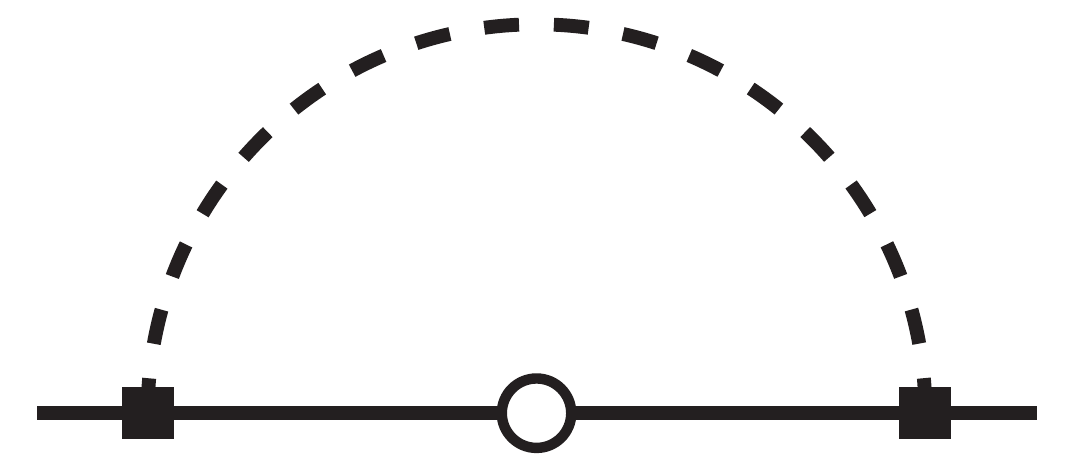}}
\subfigure[]{\label{fig:RenVertL}
\includegraphics[width=0.2\textwidth]{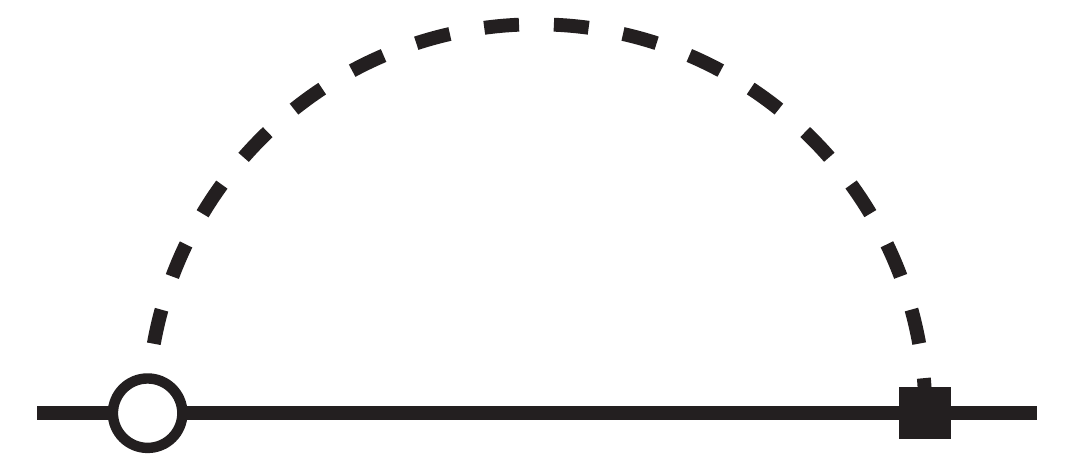}}
\subfigure[]{\label{fig:RenVertR}
\includegraphics[width=0.2\textwidth]{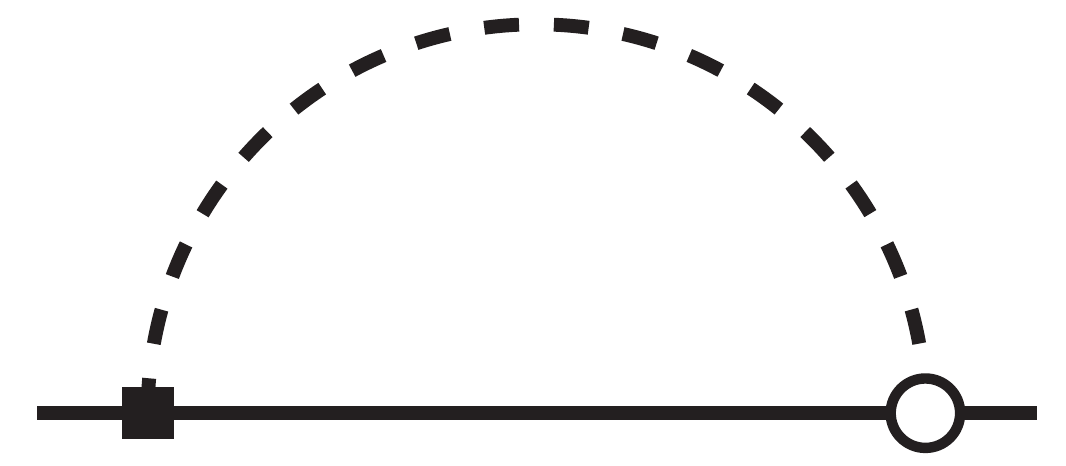}}
\subfigure[]{\label{fig:mesInsOctML}
\includegraphics[width=0.2\textwidth]{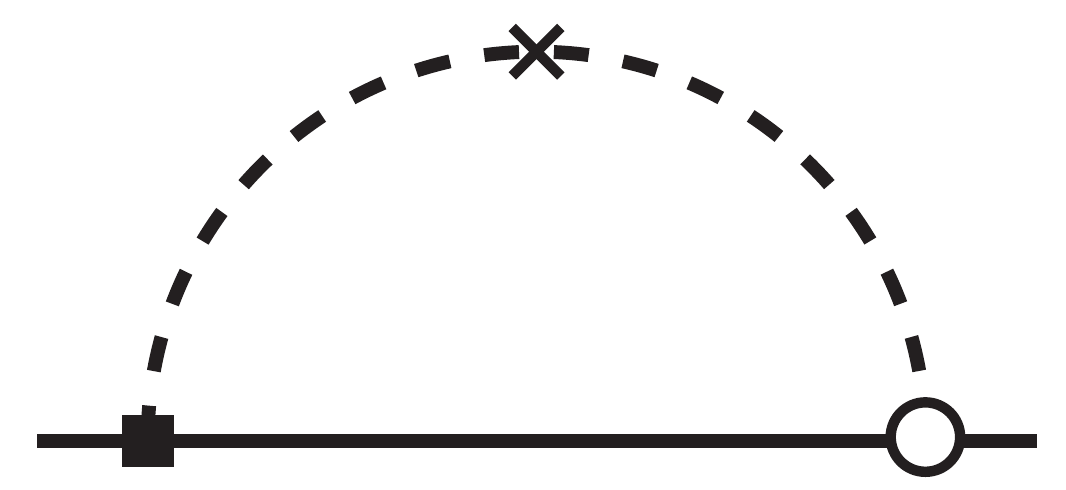}}
\subfigure[]{\label{fig:mesInsOctMR}
\includegraphics[width=0.2\textwidth]{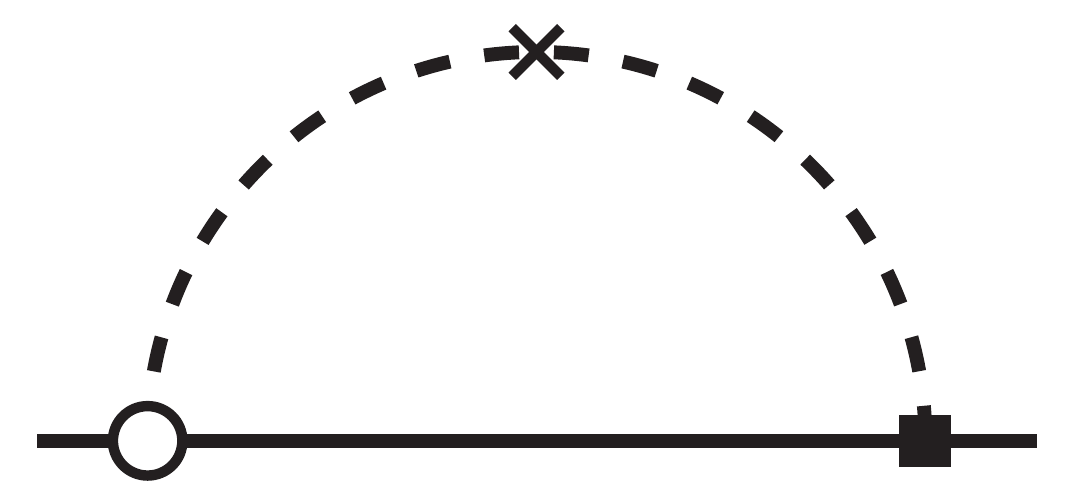}}
\subfigure[]{\label{fig:MesInsCrossLoopOctMC}
\includegraphics[width=0.2\textwidth]{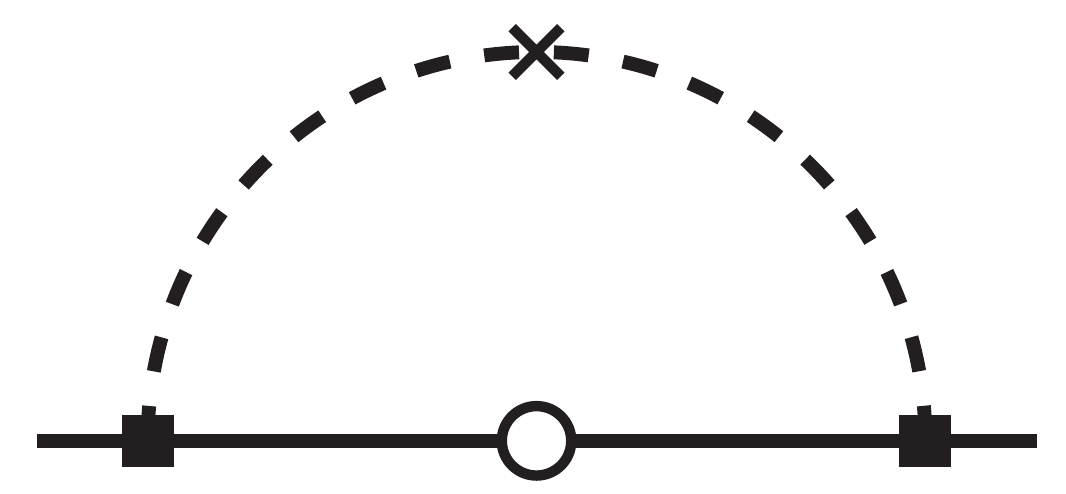}}
\subfigure[]{\label{fig:OctCrossLoopMCL}
\includegraphics[width=0.2\textwidth]{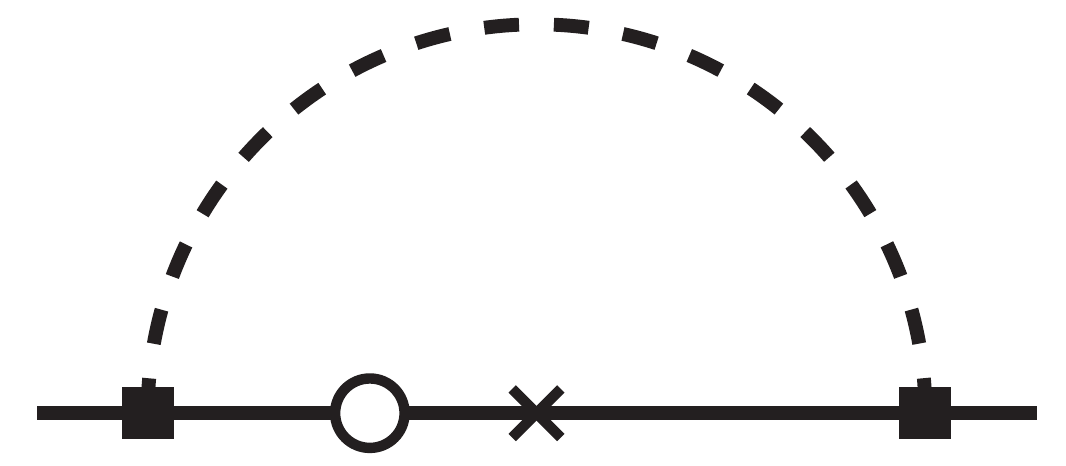}}
\subfigure[]{\label{fig:OctCrossLoopMCR}
\includegraphics[width=0.2\textwidth]{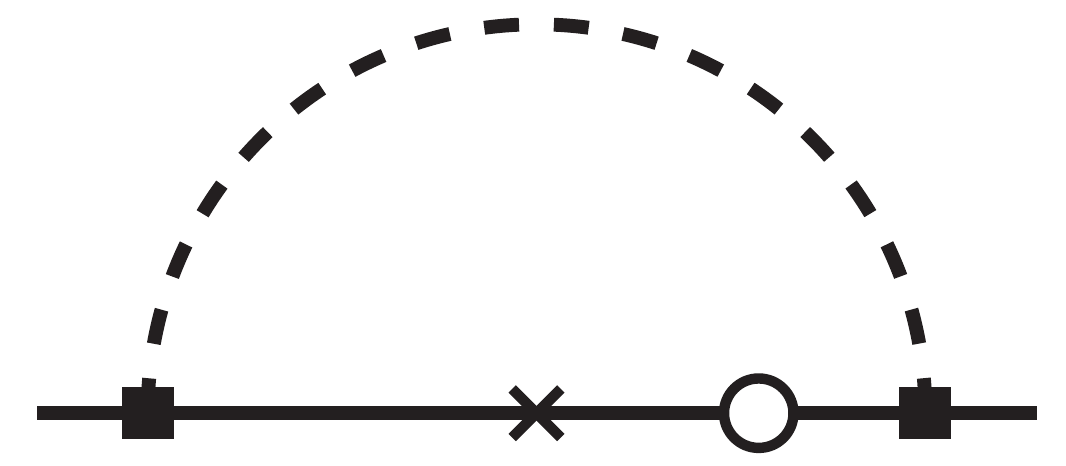}}
\subfigure[]{\label{fig:OctCrossLoopML}
\includegraphics[width=0.2\textwidth]{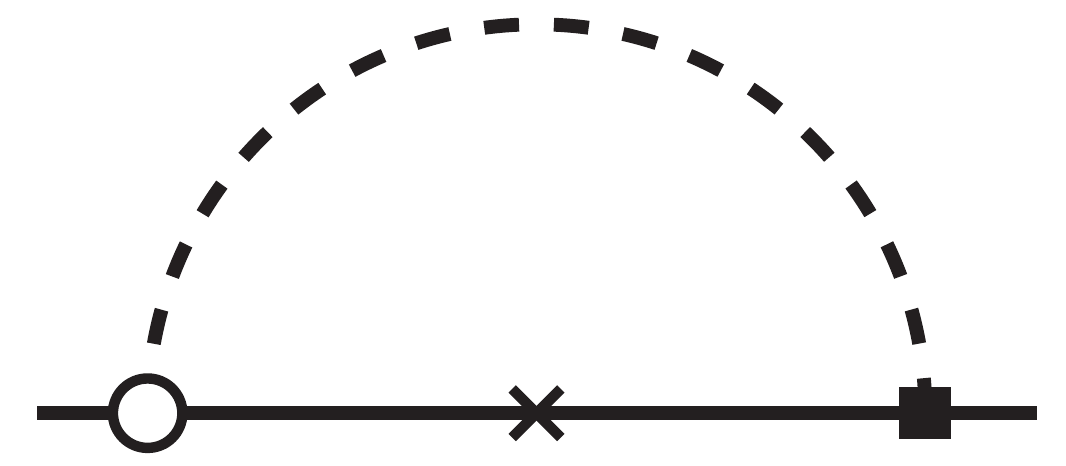}}
\subfigure[]{\label{fig:OctCrossLoopMR}
\includegraphics[width=0.2\textwidth]{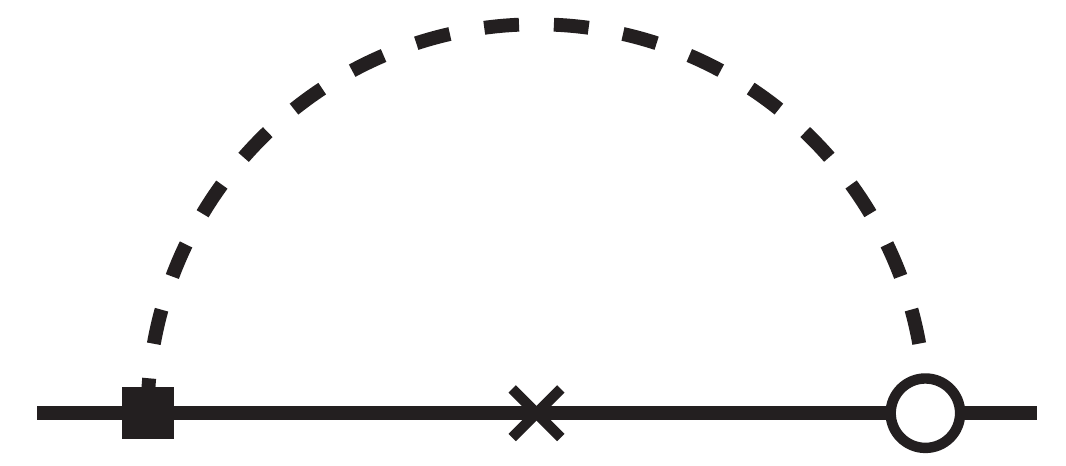}}
\subfigure[]{\label{fig:VertCrossLoopML}
\includegraphics[width=0.2\textwidth]{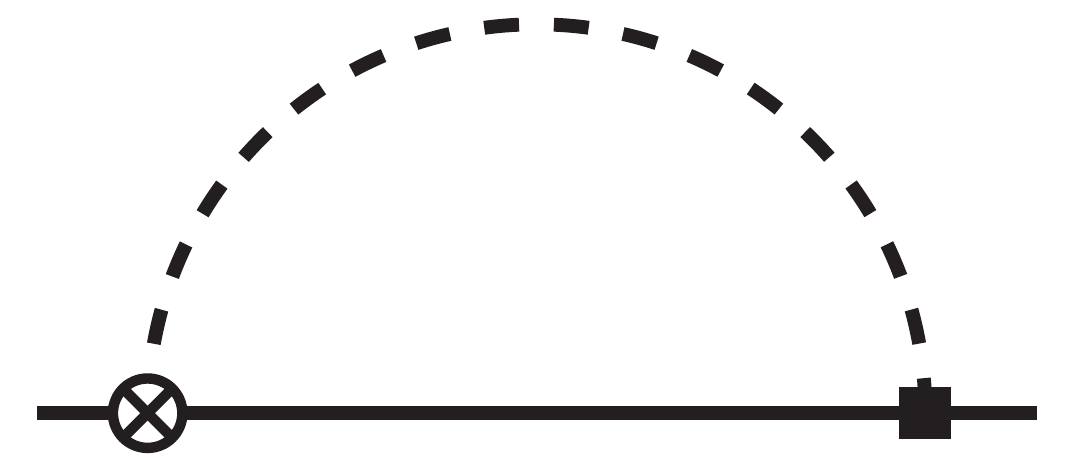}}
\subfigure[]{\label{fig:VertCrossLoopMR}
\includegraphics[width=0.2\textwidth]{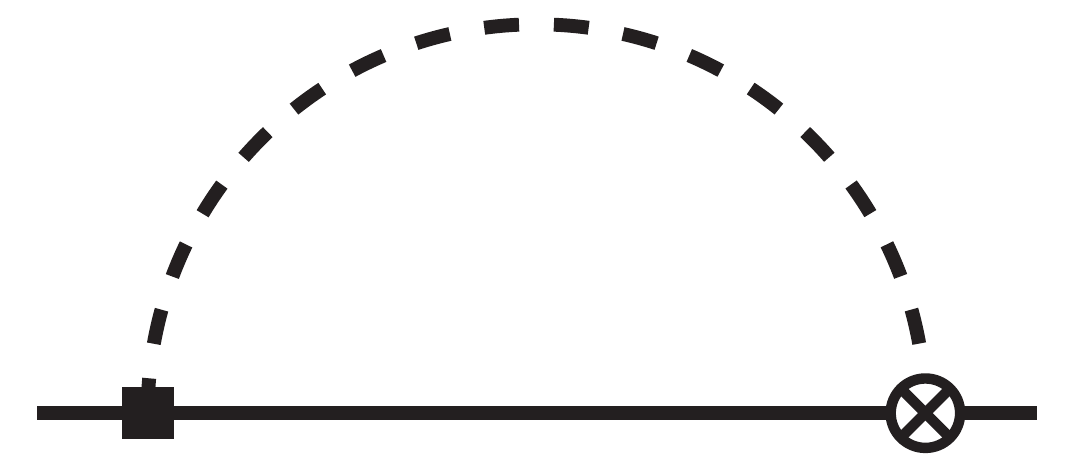}}
\caption{\label{fig:AllDiags}Loop diagrams included in the present calculation. Solid squares denote the usual strong-interaction meson-baryon vertices. The open circles and crosses represent $1/M_0$ corrections (see Eq.~\eqref{eq:1M0}) and insertions of the vector current, respectively. Single, double, and dashed lines denote octet baryons, decuplet baryons and mesons, respectively.}
\end{figure*}

The meson loop diagrams included in this calculation are shown in Fig.~\ref{fig:AllDiags}. Note that Figs.~\ref{fig:mesInsOctML} and \ref{fig:mesInsOctMR} do not contribute at $Q^2=0$ and for this reason have not been considered by other authors. The following sections give expressions for the contribution of each diagram to Eq.~\eqref{eq:Master}.

\subsubsection{Octet, Decuplet and Tadpole loops}
\label{sec:LDF}

For a transition $B_1\rightarrow B_2$, the wavefunction renormalisation diagrams (Figs.~\ref{fig:RenOct} and \ref{fig:RenDec}) give contributions 

\begin{align} \nonumber
L_\text{WR}=-\frac{1}{8\pi^2f_\pi^2}\frac{1}{2}X_R\,f^\text{SU(3)}_{1(B_1B_2)}\big(& C_{B_1B'\phi}^2I_\text{WR}(B_1,B')  \\\label{eq:WRCont}
& \left. + C_{B_2B'\phi}^2I_\text{WR}(B_2,B')  \right),
\end{align}
where
\begin{equation}
 X_R=
\begin{cases}
\frac{3}{8} & \text{for octet baryon intermediate states} \\
1 & \text{for decuplet baryon intermediate states},
\end{cases}
\end{equation}
\begin{equation}
I_\text{WR}(B,B')=\frac{4}{3}\int_0^\infty dk \left(\frac{k^4}{\omega (\omega - (M_{B}-M_{B'}))^2}\right),
\end{equation}
and a sum is implied over the repeated indices $B'$ (representing an octet or decuplet baryon intermediate state) and $\phi$ (representing the intermediate meson). Here
\begin{equation} \label{eq:omega}
\omega = \sqrt{k^2+m_\phi^2}.
\end{equation}

The baryon-baryon-meson coupling constants $C_{BB'\phi}$ are standard expressions (see, for example, Ref.~\cite{Shanahan:2014uka}) given in terms of the constants $D$, $F$ and $\mathcal{C}$. The tree-level vector current transition coefficients $f^\text{SU(3)}_{1(B_1B_2)}$ are the same as those in Eq.~\eqref{eq:Master} and are tabulated in appendix~\ref{sec:CoeffTabs}.
For our numerical results the masses of the relevant baryons $B'$ (at any values of $m_\pi$ and $m_K$), which appear in the term involving octet baryon intermediate states (Fig.~\ref{fig:RenOct}), are taken from the fits to lattice simulation results for the octet baryon masses given in Ref.~\cite{Shanahan:2012wa}.
For loops involving decuplet baryon intermediate states (Fig.~\ref{fig:RenDec}), the decuplet baryon masses (i.e., $M_{B'}$ in the above equation) are set to the value
\begin{equation}
M_\Delta = M_B^\text{SU(3)} + \Delta_{\text{phys.}},
\end{equation}
where $M_B^\text{SU(3)}$ is the SU(3)-symmetric mass of the baryon octet at the particular meson masses of relevance, and $\Delta_{\text{phys.}}=0.292$~GeV is fixed. As discussed in detail later, this choice for $\Delta$ does not affect our final results; in fact, we find entirely consistent results for $f_1$ for each transition when contributions from intermediate decuplet states are omitted entirely.

\begin{widetext}

Figures~\ref{fig:OctCrossLoop} and \ref{fig:DecCrossLoop}, with vector current insertions into the intermediate baryons in the loop diagrams, generate contributions which can be expressed as
\begin{align} \label{eq:OctInsTot}
L_\text{B}=\frac{1}{8\pi^2f_\pi^2}X_R\,C_{B_1B'\phi}\,&f^\text{SU(3)}_{1(B'B'')}\,C_{B''B_2\overbar{\phi}} I_\text{B}(B_1,B_2,B',B''),
\end{align}
where
\begin{equation*}
I_\text{B}(B_1,B_2,B',B'')=\frac{4}{3}\int_0^\infty dk\left(\frac{k^4}{\omega (\omega - (M_{B_2}-M_{B''}))(\omega - (M_{B_1}-M_{B'}))}\right),
\end{equation*}
$X_R$ is as above, sums are again implied over the indices $B'$, $B''$ and $\phi$, and the baryon masses are set as before in our numerical work.

The contributions from the loops shown in Figs.~\ref{fig:MesInsCrossLoopOct} and \ref{fig:MesInsCrossLoopDec} can be written as
\begin{equation} \label{eq:MesInsTot}
L_\text{M}=\frac{1}{8\pi^2f_\pi^2}X_R\,C_{B_1B'\phi_1}f^\text{SU(3)}_{1(\phi_1\phi_2)}C_{B'B_2\overbar{\phi_2}}I_\text{M}(B_1,B_2,B'),
\end{equation}
where
\begin{align} \nonumber
I_\text{M}(B_1,B_2,B')=-\frac{4}{3}\int_0^\infty dk\Bigg(&\frac{k^4(2\omega_2+(M_{B_1}-M_{B_2}))}{\omega_2 (\omega_1+\omega_2 + (M_{B_1}-M_{B_2})) (\omega_2-\omega_1 + (M_{B_1}-M_{B_2})) (\omega_2 - (M_{B_2}-M_{B'}))} \\ 
& + \frac{k^4(2\omega_1-(M_{B_1}-M_{B_2}))}{\omega_1 (\omega_1+\omega_2 - (M_{B_1}-M_{B_2})) (\omega_1-\omega_2 - (M_{B_1}-M_{B_2})) (\omega_1 - (M_{B_1}-M_{B'}))}\Bigg)
\end{align}
\vspace{0.4cm}
\end{widetext}
and $\omega_1$ and $\omega_2$ denote the expression given in Eq.~\eqref{eq:omega} with the masses of the intermediate mesons $m_{\phi_1}$ and $m_{\phi_2}$ replacing $m$, respectively. Again, a sum is implied over $B'$, $\phi_1$ and $\phi_2$. The $f^\text{SU(3)}_{1(\phi_1\phi_2)}$ are tabulated in appendix~\ref{sec:CoeffTabs}.

The tadpole loop diagram shown in Fig.~\ref{fig:CrossTadLoop} does not have any baryon intermediate states, so our expression reduces to that of Villadoro~\cite{Villadoro:2006nj} if dimensional regularisation is used. We can write the contribution from this loop as 
\begin{equation}
L_\text{T}\frac{1}{8\pi^2f_\pi^2}\frac{1}{2}\,C_{B_1B_2\phi V}\,I_\text{T},
\end{equation}
where
\begin{equation}
I_\text{T}= 4\int_0^\infty dk\left(\frac{k^2}{\omega}\right).
\end{equation}
The coefficients $C_{B_1B_2\phi_1\phi_2V}$ are given in appendix~\ref{sec:CoeffTabs}.

Finally, the tadpole loop shown in Fig.~\ref{fig:MesInsCrossTadLoop} gives a contribution
\begin{equation}
\frac{1}{8\pi^2f_\pi^2}\frac{1}{2}\,C_{B_1B_2\phi_1\phi_2}\,f^\text{SU(3)}_{1(\phi_1\phi_2)}\,I_\text{TM},
\end{equation}
where
\begin{widetext}
\begin{align} \nonumber
I_\text{TM}= 2\int_0^\infty dk \Bigg( &\frac{(2\omega_1+(M_{B_1}-M_{B_2}))^2}{\omega_1(\omega_1+\omega_2+(M_{B_1}-M_{B_2}))(\omega_1-\omega_2+(M_{B_1}-M_{B_2}))} \\
& - \frac{(2\omega_2-(M_{B_1}-M_{B_2}))^2}{\omega_2(\omega_1+\omega_2-(M_{B_1}-M_{B_2}))(\omega_1-\omega_2+(M_{B_1}-M_{B_2}))} \Bigg)
\end{align}
and all notation is as defined previously. The coefficients $C_{B_1B_2\phi_1\phi_2}$ are again standard coefficients which may be found, for example, in Ref.~\cite{Shanahan:2014uka}.

\end{widetext}

\subsubsection{$1/M_0$ corrections}

The terms labelled as $\mathcal{O}(1/M_0)$ corrections are relativistic corrections to the heavy-baryon formalism. They are generated by the Lagrangian pieces
\begin{subequations}
\begin{align}\label{eq:baryonProp}
\mathcal{L}_B^{(1/M_0)}= &\frac{1}{2M_0}\langle \overline{B}\left[ (v^\mu\partial_\mu)^2 - \partial^\mu\partial_\mu \right] B \rangle \\
&+ \frac{iv^\nu}{M_0} D \langle \partial_\mu\overline{B}S^\mu \{A_\nu,B\}-\overline{B}S^\mu\{A_\nu,\partial_\mu B \} \rangle \\
&+ \frac{iv^\nu}{M_0} F \langle \partial_\mu\overline{B}S^\mu [A_\nu,B]-\overline{B}S^\mu[A_\nu,\partial_\mu B ] \rangle. 
\end{align}\label{eq:1M0}
\end{subequations}
The contribution from Eq.~\eqref{eq:baryonProp} corresponds to $(1/M_0)$ corrections to the baryon propagator, which we expand as
\begin{align}\nonumber
\frac{i}{k\cdot v+i\epsilon} \rightarrow & \frac{i}{k\cdot v+i\epsilon} \\
&+ \frac{i}{k\cdot v+i\epsilon}\left(i\frac{k^2-(k\cdot v)^2}{2M_0}\right)\frac{i}{k\cdot v+i\epsilon}.
\end{align}
The remaining terms correspond to corrections to the strong-interaction vertices.
Here $M_0$ is the heavy-baryon mass scale, taken to be the octet baryon mass in the chiral limit. We use the value of $M_0$ obtained in Ref.~\cite{Shanahan:2012wa} from a chiral extrapolation of octet baryon masses. The difference between choosing the chiral-limit value or any physical baryon mass for $M_0$ is a higher-order effect. In practice we do find that the shift in our results from making such a change of convention is small. As detailed later, we add a systematic accounting for this to the quoted uncertainties in our numerical work.

Our expressions for the $\mathcal{O}(1/M_0)$ correction terms agree with Villadoro~\cite{Villadoro:2006nj} in the limit of no mass difference between the baryons, no momentum transfer, and using dimensional regularisation to evaluate the loop integral expressions. We note again that the additional diagrams in Figs.~\ref{fig:mesInsOctML} and \ref{fig:mesInsOctMR} contribute to our expressions but vanish at $Q^2=0$. We do not consider decuplet intermediate states here. We write out each contribution in turn.

The wavefunction renormalisation contribution shown in Fig.~\ref{fig:RenVertC} takes exactly the same form as Eq.~\eqref{eq:WRCont}, under the replacement $I_\text{WR}\rightarrow I_\text{WRM}$, where
\begin{equation}
I_\text{WRM}(B,B')=-\frac{4}{3 M_0}\int_0^\infty dk \left( \frac{k^6}{\omega (\omega - (M_{B}-M_{B'}))^3}\right).
\end{equation}
The total contribution from the other wavefunction renormalisation terms, depicted in Figs.~\ref{fig:RenVertL} and \ref{fig:RenVertR}, also takes exactly the same form as Eq.~\eqref{eq:WRCont}, under the replacement $I_\text{WR}\rightarrow I_\text{WRMV}$, where
\begin{equation}
I_\text{WRMV}(B,B')=-\frac{4}{3 M_0} \int_0^\infty dk\left( \frac{k^4}{(\omega - (M_{B}-M_{B'}))^2}\right).
\end{equation}

\begin{widetext}

The total contribution from the diagrams shown in Figs.~\ref{fig:mesInsOctML} and \ref{fig:mesInsOctMR} can be expressed in the form of Eq.~\eqref{eq:MesInsTot}, under the replacement $I_\text{M} \rightarrow I_\text{MMV}$, where
\begin{align} \nonumber
I_\text{MMV}(B_1,B_2,B')=\\\nonumber
-\frac{4}{3}\frac{1}{4M_0}\int_0^\infty dk\Bigg(&\frac{k^4(2\omega_2+(M_{B_1}-M_{B_2}))^2}{\omega_2 (\omega_2+\omega_1 + (M_{B_1}-M_{B_2})) (\omega_2-\omega_1 + (M_{B_1}-M_{B_2})) (\omega_2 - (M_{B_2}-M_{B'}))} \\ 
& + \frac{k^4(2\omega_1-(M_{B_1}-M_{B_2}))^2}{\omega_1 (\omega_1+\omega_2 - (M_{B_1}-M_{B_2})) (\omega_1-\omega_2 - (M_{B_1}-M_{B_2})) (\omega_1 - (M_{B_1}-M_{B'}))}\Bigg).
\end{align}
Similarly, the contribution from Fig.~\ref{fig:MesInsCrossLoopOctMC} can be expressed in the form of Eq.~\eqref{eq:MesInsTot}, under the replacement $I_\text{M} \rightarrow I_\text{MM}$, where
\begin{align} \nonumber
I_\text{MM}(B_1,B_2,B')=\\\nonumber
\frac{4}{3}\frac{1}{2M_0}\int_0^\infty dk\Bigg(&\frac{k^6(2\omega_2+(M_{B_1}-M_{B_2}))}{\omega_2 (\omega_2+\omega_1 + (M_{B_1}-M_{B_2})) (\omega_2-\omega_1 + (M_{B_1}-M_{B_2})) (\omega_2 - (M_{B_2}-M_{B'}))^2} \\ 
& + \frac{k^6(2\omega_1-(M_{B_1}-M_{B_2}))}{\omega_1 (\omega_1+\omega_2 - (M_{B_1}-M_{B_2})) (\omega_1-\omega_2 - (M_{B_1}-M_{B_2})) (\omega_1 - (M_{B_1}-M_{B'}))^2}\Bigg).
\end{align}

The total contribution from Figs.~\ref{fig:OctCrossLoopMCL} and \ref{fig:OctCrossLoopMCR} can be expressed in the same form as Eq.~\eqref{eq:OctInsTot}, under the replacement $I_\text{B}\rightarrow I_\text{BM}$, where
\begin{align}\nonumber
I_\text{BM}(B_1,B_2,B',B'')=-\frac{2}{3 M_0} \int_0^\infty dk\Bigg( & \frac{k^6}{\omega (\omega - (M_{B_1}-M_{B'}))^2(\omega - (M_{B_2}-M_{B''}))} \\ & + \frac{k^6}{\omega (\omega - (M_{B_1}-M_{B'}))(\omega - (M_{B_2}-M_{B''}))^2}\Bigg).
\end{align}
The contribution from Figs.~\ref{fig:OctCrossLoopML} and \ref{fig:OctCrossLoopMR} can also be expressed in the same form as Eq.~\eqref{eq:OctInsTot}, under the replacement $I_\text{B}\rightarrow I_\text{BMV}$, where
\begin{equation}
I_\text{BMV}(B_1,B_2,B',B'')=-\frac{4}{3 M_0}\int_0^\infty dk \left( \frac{k^4}{(\omega - (M_{B_1}-M_{B'})) (\omega - (M_{B_2}-M_{B''}))}\right).
\end{equation}
Finally, the contribution from Figs.~\ref{fig:VertCrossLoopML} and \ref{fig:VertCrossLoopMR} can be expressed as
\begin{equation}
L_\text{V}^\text{oct}=\frac{1}{8\pi^2f_\pi^2}X_R\,\frac{1}{2}\,\left(C_{B_1B'\phi V}C_{B'B_2\overline{\phi}}I_\text{BMV}(0,B_2,0,B')-C_{B_1B'\phi}C_{B'B_2\overline{\phi}V}I_\text{M}(B_1,0,B',0)\right).
\end{equation}

\end{widetext}

\subsubsection{Master formula}

The master formula for the chiral extrapolation, Eq.~\eqref{eq:Master}, can now be re-written in terms of the explicit expressions for the loop diagram contributions:

\begin{align}\nonumber
f_1^{B_1B_2}&(Q^2=(M_B-M_{B'})^2) \\\nonumber
= & f_1^\text{SU(3)} + C_{B_1B_2}Q^2 + L_\text{T} \\ \nonumber
& + L_\text{WR}^\text{oct} + L_\text{M}^\text{oct}  + L_\text{WR}^\text{dec}+L_\text{M}^\text{dec}\\\label{eq:Master2}
& + +L_\text{WRM}^\text{oct}+L_\text{WRMV}^\text{oct}+L_\text{BM}^\text{oct}+L_\text{BMV}^\text{oct}+L_\text{V}^\text{oct}.
\end{align}
In practice, we evaluate each loop integral expression within the finite-range regularisation (FRR) scheme. This choice of regularization procedure is discussed in detail in Refs.~\cite{Leinweber:2003dg,Young:2002cj,Young:2002ib}. In short, the inclusion of a finite cutoff into the loop integrands effectively resums the chiral expansion in a way that suppresses the loop contributions at large meson masses. This enforces the physical expectation, based on the finite size of the baryon, that meson emission and absorption processes are suppressed for large momenta~\cite{Thomas:2002sj}. For the case of the octet baryon masses, FRR appears to offer markedly improved convergence properties of the (traditionally poorly convergent~\cite{Donoghue:1998bs,Borasoy:2002jv}) SU(3) chiral expansion, and this scheme consistently provides robust fits to lattice data at leading or next-to-leading order. Nevertheless, one could calculate the size of higher order corrections to confirm that these contributions are small as expected. Changing to dimensionally regularized integral expressions requires simple substitutions; details are given in~\cite{Borasoy:2002jv}.

For our numerical analysis we choose a dipole regulator $u(k)=\left(\frac{\Lambda^2}{\Lambda^2+k^2}\right)^2$ with a regulator mass $\Lambda=0.8\pm 0.2$~GeV. This form is suggested by a comparison of the nucleon's axial and induced pseudoscalar form factors and the choice of $\Lambda$ is supported by a lattice analysis of nucleon magnetic moments~\cite{Guichon:1982zk,Hall:2012pk}. 
We note that different regulator forms, for example monopole, Gaussian or sharp cutoff yield extrapolated results which are consistent within the quoted uncertainties.

The most recent comparable effective field theory studies are those by Villadoro~\cite{Villadoro:2006nj}, which is a heavy-baryon calculation (with dimensional regularisation) with both decuplet intermediate states and relativistic ($1/M_0$) corrections considered, Lacour~\cite{Lacour:2007wm}, which is a covariant calculation that neglects the decuplet, and Geng~\cite{Geng:2014efa}, which is a covariant calculation with decuplet degrees of freedom.
The primary difference between these works and ours, other than our non-zero $Q^2$, is that we have more carefully accounted for the meson-mass dependence of the mass-splittings among the octet baryons. While other authors have either fixed the mass differences between members of the baryon octet (which appear in the loop integral expressions) absolutely, or varied them linearly, we have used the chiral extrapolation of octet baryon masses which was presented in Ref.~\cite{Shanahan:2012wa} to give these mass differences as a function of meson masses. 
Furthermore, while Villadoro Taylor-expands in the mass differences between the octet baryons, we include them explicitly (i.e., resumming Villadoro's expansion to all orders). We also include the octet baryon mass splittings in the $1/M_0$ correction diagrams (which account for relativistic corrections to the heavy-baryon approach), which was not done previously.
Our expressions match those from Villadoro~\cite{Villadoro:2006nj}, which is the most closely related calculation, in the relevant limits.

\section{Lattice simulation}

We apply the formalism developed in the previous sections to new lattice simulation results for the $\Sigma\rightarrow N$ and $\Xi \rightarrow \Sigma$ transition vector form factors. Preliminary results appeared in Refs.~\cite{Cooke:2013qqa,Cooke:2012xv}.

We use gauge field configurations with 2+1 flavours of nonperturbatively $\mathcal{O}(a)$-improved Wilson fermions. The clover action consists of the tree-level Symanzik improved gluon action together with a mild `stout' smeared fermion action~\cite{Bietenholz2010436}. We use a single lattice volume, $L^3T = 32^3\times 64$, with $\beta=5.5$, corresponding to the lattice scale $a = 0.074(2)$~fm (set using various singlet quantities~\cite{Horsley:2013wqa,Bietenholz:2011qq,Bietenholz2010436}). Details are given in Table~\ref{tab:rawLatt}. The simulations correspond to two distinct fixed-singlet-mass trajectories in $m_\pi$--$m_K$ space.

The vector form factor $f_1$ is defined in terms of the matrix elements of the charged strangeness-changing ($s\rightarrow u$) weak vector current $V^\mu = \overline{u}\gamma^\mu s$. In Euclidean space, defining $q^2=-Q^2$,
\begin{align}
\begin{split}
\langle B_2 | V_\mu | B_1 \rangle = \overline{B}_2(p_2)\Bigg[ \gamma_\mu f_1\left(q^2\right)+\frac{\sigma_{\mu\nu}q_\nu}{M_{B_1}+M_{B_2}}f_2\left(q^2\right) \\
 + \frac{iq_\mu}{M_{B_1}+M_{B_2}}f_3\left(q^2\right) \Bigg] B_1(p_1).
\end{split}
\end{align}
While here we are interested in the vector form factor $f_1$ only, what is accessible on the lattice at $q^2_\text{max}=(M_{B_1}-M_{B_2})^2$ is in fact the `scalar form factor':
\begin{equation}\label{eq:f0eq}
f_0(q^2)=f_1(q^2)+\frac{q^2}{M_{B_1}^2-M_{B_2}^2}f_3(q^2).
\end{equation}
This quantity can be obtained with high precision from the ratio
\begin{equation}
R(t,t') = \frac{G_4^{B_1B_2}(t',t;\vec{0},\vec{0})G_4^{B_2B_1}(t',t;\vec{0},\vec{0})}{G_4^{B_1B_1}(t',t;\vec{0},\vec{0})G_4^{B_2B_2}(t',t;\vec{0},\vec{0})} 
\end{equation}
which tends to $\left| f_0(q^2_\text{max})\right|^2$ in the limit $t,(t'-t)\rightarrow \infty$. Here, for example, $G_4^{B_1B_2}(t',t;\vec{0},\vec{0})$ is the zero three-momentum lattice three-point function of the fourth component of the vector current $V^4$, inserted at time $t$ between the source baryon $B_1$ located at time $t = 0$ and the sink baryon $B_2$ at time $t'$. In the SU(3) flavour symmetric limit, $R(t',t) = 1$; any deviations from unity are purely due to symmetry-breaking effects.

\begin{table*}
\centering
\begin{tabular}{ccccccc | cc |cc }\hline 
&& & & & & & \multicolumn{2}{c}{$f_1^{\Sigma^- n}$} & \multicolumn{2}{c}{$f_1^{\Xi^0\Sigma^+}$}\\
 &$\kappa_0$ & $\kappa_l$ & $\kappa_s$ & $m_\pi$~(MeV) & $m_K$~(MeV) & $m_\pi L$ & $q^2$~(GeV$^2$) & lattice value & $q^2$~(GeV$^2$) & lattice value \\\hline
 1&0.120900 & 0.121040 &0.120620& 360& 505& 4.3 & 0.0017 & -1.000(3) & 0.0006 & 0.990(3) \\
 2 && 0.121095 &0.120512& 310& 520& 3.7 & 0.0042 & -0.999(7) & 0.0005 & 0.985(5) \\\hline
 3 & 0.120950 & 0.121040 &0.120770& 330& 435& 4.0 & 0.0006 & -1.002(4) & 0.0002 & 0.994(3) \\ \hline
\end{tabular}
\caption{\label{tab:rawLatt}
Details of the lattice simulation parameters and raw lattice simulation results for the hyperon vector transition form factors at $\vec{q}=0$ with fixed zero sink momentum. The parameter $\kappa_0$ denotes the value of $\kappa_l = \kappa_s$ at the SU(3)-symmetric point, $\beta=5.5$ corresponding to $a = 0.074(2)$~fm, and $L^3\times T = 32^3\times 64$.}
\end{table*}

\begin{figure}
\centering
\includegraphics[width=0.48\textwidth]{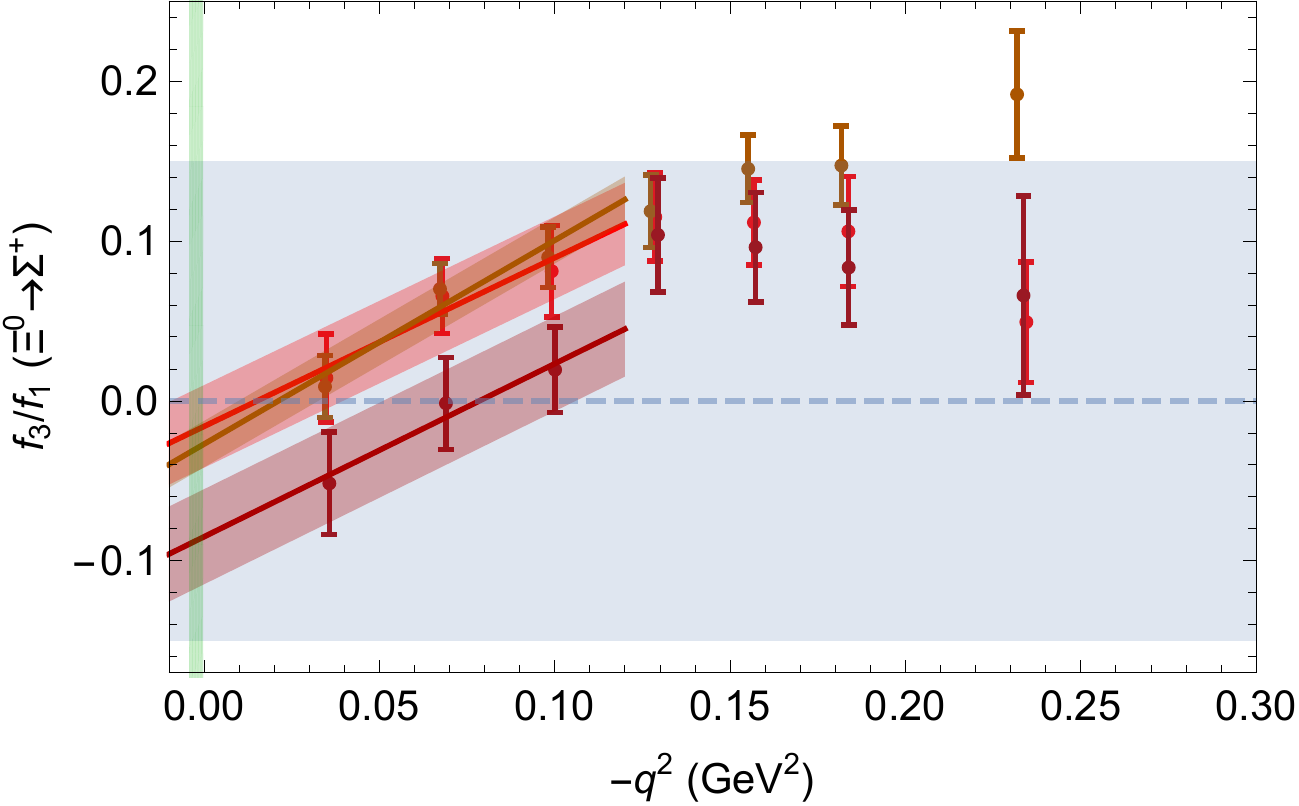}
\includegraphics[width=0.48\textwidth]{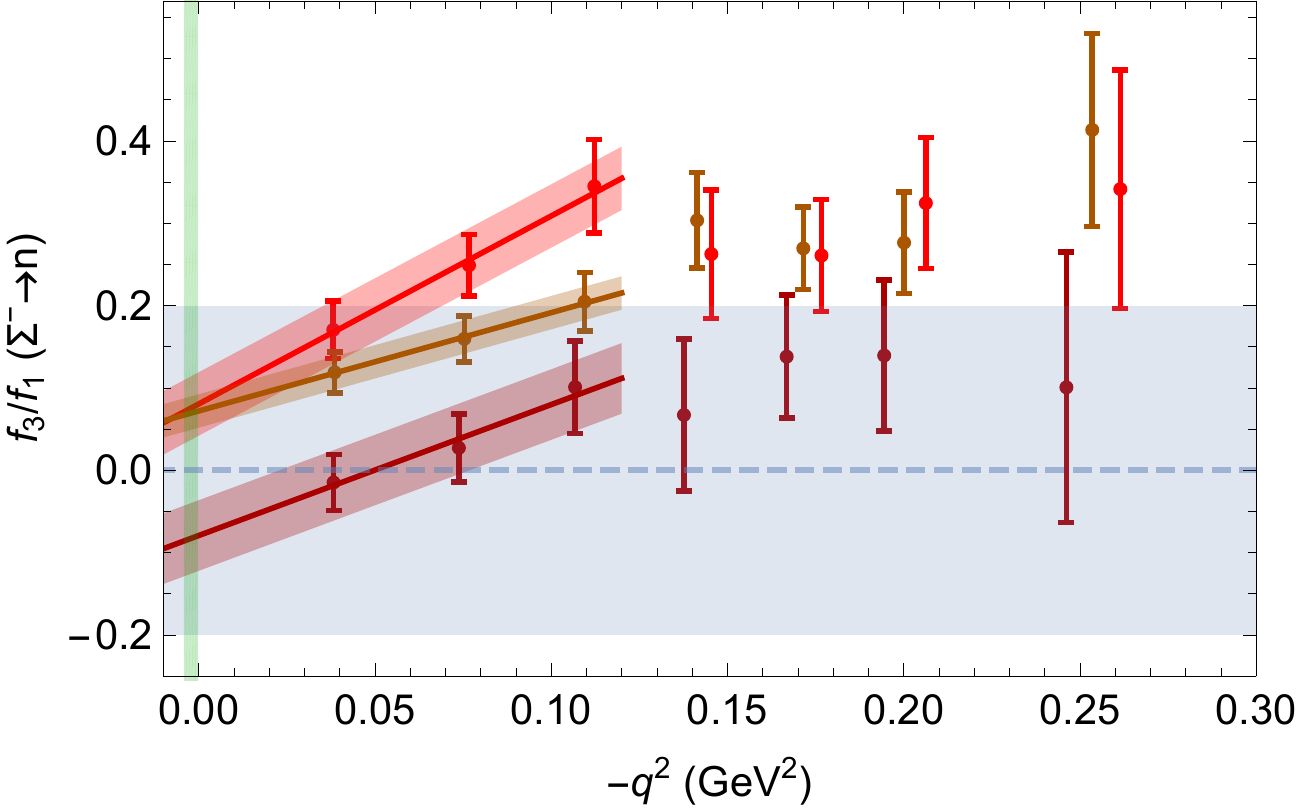}
\caption{\label{fig:vffratio}Ratio of $f_3$ to $f_1$ for the two transitions for which we produce lattice simulation results. The three sets of data (brown, red, purple) on each plot correspond to the three sets of pseudoscalar masses (labelled 1, 2, 3) in Table~\ref{tab:rawLatt}. The vertical green band shows the range of $q^2$ values for which we wish to estimate $f_3/f_1$, while the horizontal blue band shows our bounds on this ratio at the $q^2$ values of interest based on the linear extrapolations of the simulation results which are shown as brown, red and purple bands.}
\end{figure}

Away from $q^2_\text{max}=(M_{B_1}-M_{B_2})^2$, it is possible to determine both $f_1$ and $f_3$ independently~\cite{Guadagnoli200763,PhysRevD.79.074508,PhysRevD.86.114502}. The ratio of these two quantities, at the pseudoscalar masses of our simulations, is shown in Fig.~\ref{fig:vffratio}. At the relevant values of $q^2_\text{max}=(M_{B_1}-M_{B_2})^2$, indicated by the green vertical band on the figure, it is clear that $|f_3/f_1|<\{0.15,0.2\}$ (where the two numbers are for the $\Xi\rightarrow\Sigma$ and $\Sigma\rightarrow N$ transitions respectively) is a conservative bound.
This is supported by both quenched lattice simulations~\cite{Guadagnoli200763,PhysRevD.79.074508} and quark models~\cite{Dahiya:2008xj} which find small values for that ratio over a range of values of $q^2$. 
Taking these bounds, scaled by the appropriate kinematic factors, the contribution from $f_3$ to $f_0$ is then for each set of simulation pseudoscalar masses negligible compared with the statistical uncertainties of the calculation. That is, to the statistical precision of our simulation, $f_0$ and $f_1$ are the same, so we take our simulation value of $f_0$ as the result for the vector form factor.

We access the strangeness-changing transitions involving only the outer-ring octet baryons, i.e., $\Sigma \rightarrow  N$ and $\Xi\rightarrow  \Sigma$.
Simulation results are given in Table~\ref{tab:rawLatt}.
We note that the simulations suffer no systematic uncertainties from omitted disconnected loops because these terms cannot contribute to the transitions.

\section{Fits to the lattice results}

\begin{figure}
\centering
\includegraphics[width=0.48\textwidth]{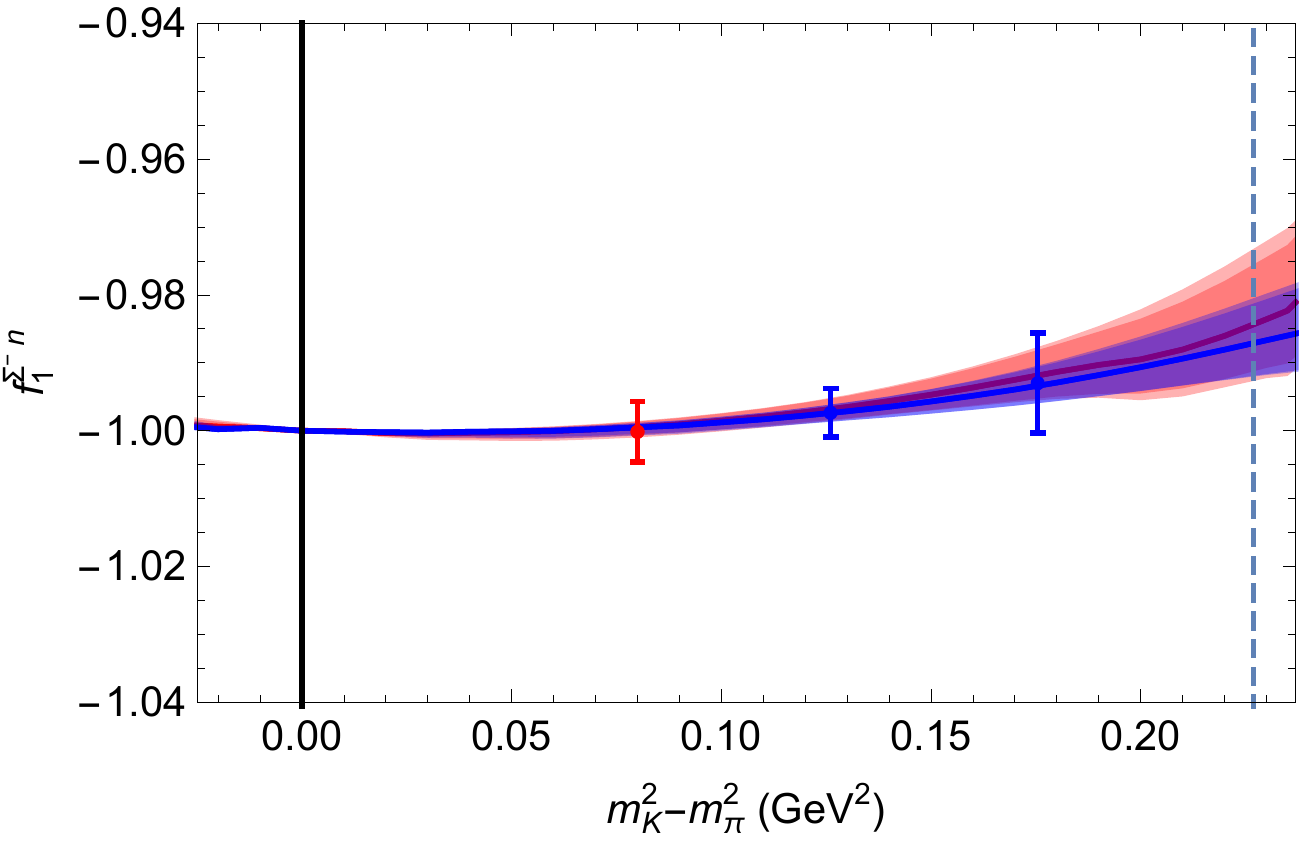}\\
\includegraphics[width=0.48\textwidth]{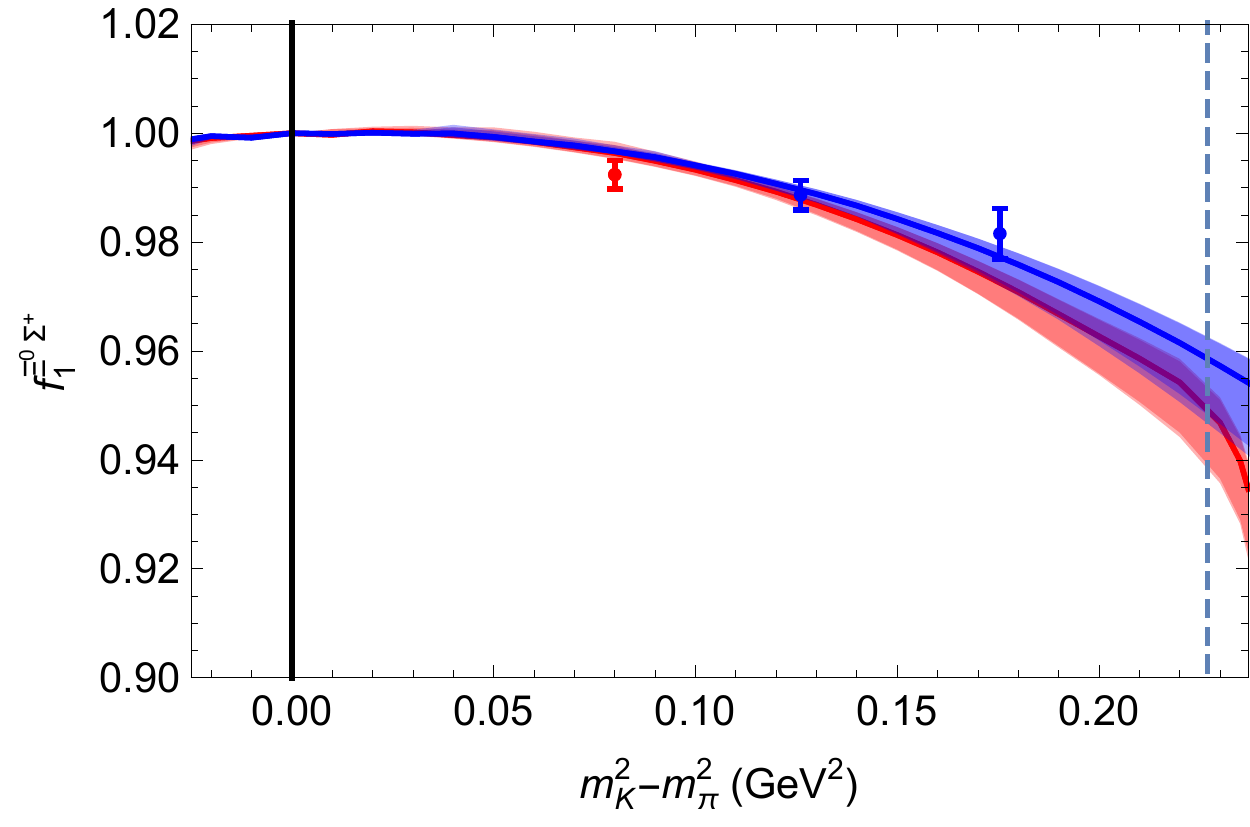}
\caption{\label{fig:vffFits}Vector form factors as a function of the SU(3)-breaking parameter $m_K^2-m_\pi^2$. The two (blue and red) trajectories on each plot correspond to the two different constant-singlet-mass lines (trajectories \{1,2\} and 3 in Table~\ref{tab:rawLatt}, respectively) on which the simulations lie. The error bands are as described in the text. The vertial dashed line denotes the physical pseudoscalar mass point.}
\end{figure}

Before performing a chiral extrapolation of this lattice data to the physical pseudoscalar masses using the formalism presented in the previous sections, we correct for the finite simulation lattice volume using the difference between infinite-volume integrals and finite-volume sums for the loop integral expressions in the chiral expansion. The procedure used here follows Ref.~\cite{Hall:2013oga}. There are no free low-energy constants in the finite-volume correction expressions; they depend only on $F$, $D$ and $\mathcal{C}$. We fix these constants to their SU(6) values, namely  $D + F = g_A = 1.27$, $F=2/3D$, $\mathcal{C} = 2D$. We also set the pion decay constant in the chiral limit to $f_\pi = 0.0871$~GeV~\cite{Amoros:2001cp}. 
The magnitude of the finite volume corrections is in the range 0.1--0.6\% of the relevant form factors for each transition and set of pseudoscalar masses considered. These shifts, while small, are significant compared with the percent-level SU(3)-breaking effects of interest here. In all cases the finite volume corrections act to enhance the size of SU(3)-breaking. To account for any model-dependence in our estimate of the corrections, we allow $F$, $D$ and $\mathcal{C}$ to each vary by 20\% about their SU(6) values and add the resulting shift in the finite volume corrections in quadrature to the statistical uncertainty on the lattice simulation results. These additional uncertainties have a magnitude approximately 25\% as large as the finite-volume corrections themselves. To examine the effect of these corrections on our final results, we perform the entire fit and analysis described below to the uncorrected data set; a comparison is shown in the results section. 
The final extrapolated results with and without finite-volume corrections are consistent at 1-sigma for all transitions.

After correcting the simulation results to infinite volume, we fit the free low-energy constants $b_D$ and $b_F$ to the lattice simulation results to perform a chiral extrapolation to the physical pseudoscalar masses. For each choice of the constants $F$, $D$ and $\mathcal{C}$, which are varied within 20\% about their SU(6) values, and the FRR dipole regulator mass $\Lambda$, which is allowed to vary in the range 0.6--1~GeV, $b_D$ and $b_F$ are fit to the numerical simulation results. The quality of fit is illustrated in Fig.~\ref{fig:vffFits}. The central lines of each band show the fit using the central values of $F$, $D$, $\mathcal{C}$ and $\Lambda$. The width of the shaded regions is determined from the statistical 1$\sigma$ variation generated by the fit of $b_D$ and $b_F$ with the central values of $F$, $D$, $\mathcal{C}$ and $\Lambda$, added in quadrature with the shift in central values obtained when fitting $b_D$ and $b_F$ with the range of choices of $F$, $D$, $\mathcal{C}$ and $\Lambda$. For our final results which are presented in the next section, we also add an uncertainty calculated by allowing the value of the heavy-baryon mass scale $M_0$ to vary between the chiral-limit value and the average octet baryon mass at the physical point. 

Finally, we comment that at the physical pseudoscalar masses we do find nonanalytic corrections to the SU(3)-breaking beyond the Ademollo-Gatto terms quadratic in the strange-nonstrange quark mass difference. These are a consequence of the opening of the meson-baryon decay channels as the physical degree of SU(3)-breaking is approached. For instance, the $\Sigma\rightarrow N$ amplitude exhibits a cusp at the point where the imaginary part associated with the decay $\Sigma\rightarrow N \pi$ appears. The numerical influence of these terms beyond the standard $(m_s-m_l)^2$ effects tend to be much smaller than the statistical uncertainties of this calculation.

\section{Figures and Results}

\begin{table}
\centering
\begin{tabular}{cccd{2.8}c}\hline
$B_1$ & $B_2$ & $f_1^\text{SU(3)}$ & \multicolumn{1}{c}{$\left(f_1/f_1^\text{SU(3)}-1\right)\times 100$~} & $Q^2$~(GeV$^2$)  \\ \hline
$\Sigma$ & $N$ & -1 & -1.5(11)\substack{+5 \\ -8}(2) & -0.065 \\
$\Xi$ & $\Sigma$ & 1 & -4.8(7)\substack{+3 \\ -10}(4) & -0.016 \\
$\Lambda$ & $p$ & $-\sqrt{\frac{3}{2}}$ &-4.5(7)\substack{+5 \\ -16}(3) & -0.031 \\
$\Xi^-$ & $\Lambda$ & $\sqrt{\frac{3}{2}}$ &  -5.4(7)\substack{+12 \\ -32}(0) & -0.041 \\\hline
\end{tabular}
\caption{\label{tab:Results}Results for $f_1(Q^2=-(M_{B_1}-M_{B_2})^2)$ at the physical pseudoscalar masses and extrapolated to infinite volume. The fourth column shows the SU(3)-breaking corrections as a percentage. The first uncertainty is statistical and also includes the uncertainty in the finite-volume corrections. The second uncertainty allows for variation of the low-energy constants $D$, $F$ and $\mathcal{C}$ and the FRR dipole regulator mass $\Lambda$, and the third the variation of $M_0$, as described in the text.
}
\end{table}

\begin{figure}
\includegraphics[width=0.49\textwidth]{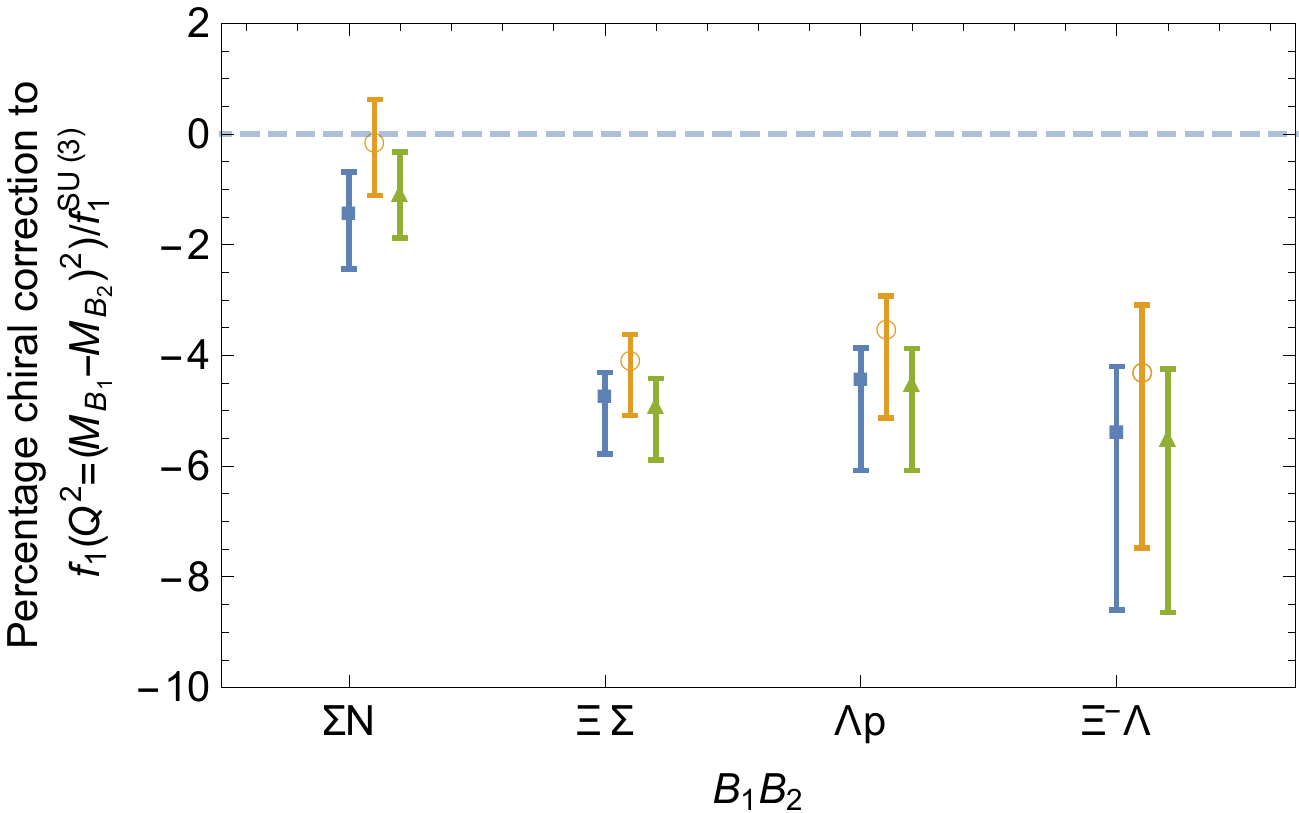}
\caption{\label{fig:PercentFig}Percentage SU(3)-breaking in $f_1$ at $Q^2_\text{max}=-(M_{B_1}-M_{B_2})^2$ for each of the strangeness-changing transitions considered here. The three sets of data points correspond to our full results (blue solid squares), as well as the results applying an identical analysis but omitting the finite volume corrections (orange open circles) or omitting decuplet intermediate states (green solid triangles).}
\end{figure}

\begin{figure*}[]
\includegraphics[width=0.67\textwidth]{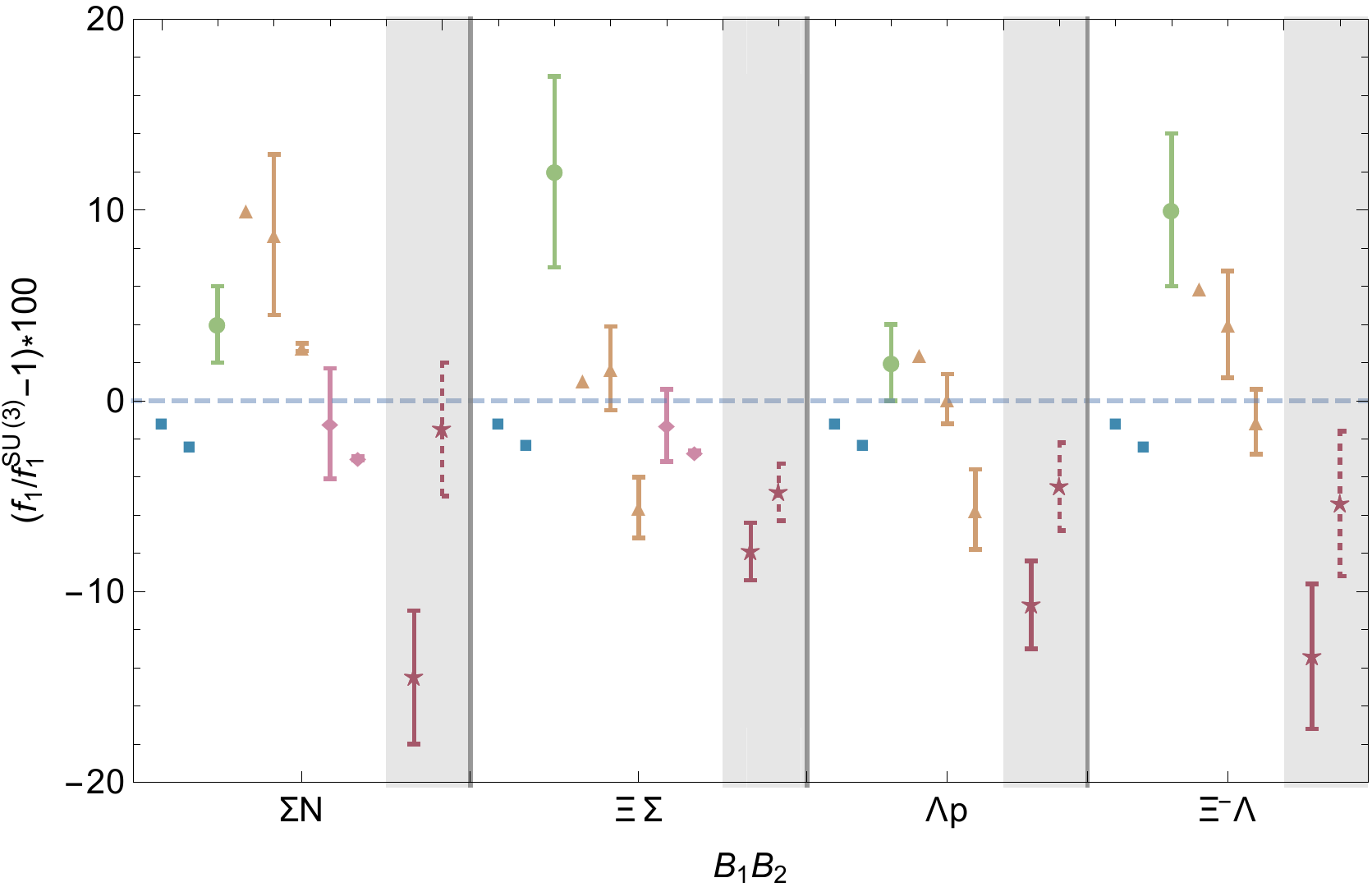}
\caption{\label{fig:summaryFig} Comparison of percentage SU(3)-breaking in $f_1$ determined in this work, highlighted by the shaded bands, with that of other calculations. The error bands for our results are those given in Table~\ref{tab:Results} combined in quadrature. Blue squares, green circles and orange triangles denote results of quark model~\cite{Donoghue:1986th,Schlumpf:1994fb}, $1/N_c$ expansion~\cite{PhysRevD.58.094028} and chiral perturbation theory~\cite{Anderson:1993as,Geng:2009ik,Lacour:2007wm} approaches respectively, while the pink diamonds show results from lattice QCD~\cite{Guadagnoli200763,PhysRevD.86.114502}. The red stars show the results of this work at $Q^2=0$ (solid line), where we have corrected from $\vec{q}=0$ to $Q^2=0$ using the dipole form given in Eq.~\eqref{eq:dipole}, and at $Q^2=-(M_B-M_{B'})^2$ (dotted line).}
\end{figure*}

The results for the vector transition form factors, after infinite-volume and chiral extrapolation, are summarized in Table~\ref{tab:Results}. 
Recall that these results are at $Q^2=-(M_{B_1}-M_{B_2})^2$ (i.e., corresponding to $\vec{q}=0$ in our lattice simulations with fixed zero sink momentum), with the physical values of the baryon masses $B_1$ and $B_2$, instead of at $Q^2=0$ as is standard. Moving to $Q^2=0$ would reduce the magnitude of each form factor, i.e., increase the SU(3)-breaking effect in each case (as will be shown explicitly later). As described in the previous section, the quoted uncertainties allow for 20\% variation of the low-energy constants $D$, $F$ and $\mathcal{C}$ from their SU(6) values, and for the FRR dipole regulator mass $\Lambda$ to vary in the range 0.6-1~GeV. Furthermore, we allow $M_0$, the heavy-baryon mass scale used to account for leading relativistic (or kinematic) corrections, to vary between the chiral-limit value and the average octet baryon mass at the physical point. We also account for uncertainties in the finite-volume corrections as described in the previous section.

Figure~\ref{fig:PercentFig} shows the results from Table~\ref{tab:Results} graphically, alongside the results obtained using an identical analysis but omitting either finite-volume corrections or contributions from decuplet baryon intermediate states. Clearly, all results are stable under these changes. Previous pure--effective-field-theory calculations of these quantities (e.g., Ref.~\cite{Geng:2014efa}) have typically been very sensitive to decuplet baryon effects. We attribute this difference primarily to our use of the FRR scheme.

Following the work in Refs.~\cite{Shanahan:2015caa,Shanahan:2012wa,Shanahan:2013vla}, we are also able to use the chiral extrapolation formalism to determine the effect of a non-zero light quark mass difference ($m_u\ne m_d$) on our results. As we find such charge-symmetry violating effects to be one to two orders of magnitude smaller than the SU(3)-breaking effects, we neglect these differences. Explicitly, we find the difference in the quantity $(f_1/f_1^\text{SU(3)}-1)\times 100$ for $\Sigma^-\rightarrow n$ and $\Sigma^0 \rightarrow p$ and also $\Xi^0 \rightarrow \Sigma^+$ and $\Xi^- \rightarrow \Sigma^0$ to be in the range 0.03--0.04, which is an order of magnitude smaller than the statistical uncertainties of our calculation.

Finally, to estimate the magnitude of the effect caused by the non-zero values of $Q^2$ used in our analysis, we have corrected from $Q^2=-(M_{B_1}-M_{B_2})^2$ to $Q^2=0$ using the standard dipole parameterisation which is used to fit experimental results~\cite{FloresMendieta:2004sk}:
\begin{equation}\label{eq:dipole}
f_1(Q^2) = \frac{f_1(0)}{(1+Q^2/M_V^2)^2},
\end{equation}
where $M_V=0.97$~GeV is chosen, generally universally across the baryon octet, for strangeness-changing (and 0.84~GeV for strangeness-conserving) decays~\cite{Ratcliffe:2004jt}. 
These numbers may be more directly compared with the results of previous analyses as shown in Fig.~\ref{fig:summaryFig}. It is clear that the naive extrapolation in $Q^2$ by Eq.~\eqref{eq:dipole} causes a significant enhancement of the SU(3)-breaking in our results, particularly for the $\Sigma\rightarrow N$ transition where in our calculation the value of $Q^2$ is the largest. We emphasize that our numerical results are presented in Table~\ref{tab:Results} and obtained at non-zero values of $Q^2$; the $Q^2=0$ results are merely shown to facilitate comparison with other work and are obtained using Eq.~\eqref{eq:dipole} with no attempt to quantify the model-dependence of the extrapolation.

It is clear from Fig.~\ref{fig:summaryFig} that quark models in general predict negative corrections from SU(3)-breaking~\cite{Donoghue:1986th,Schlumpf:1994fb} to the vector form factors. This agrees in sign with our results although we predict more significant SU(3)-breaking effects in all channels, particularly after extrapolation to $Q^2=0$. Previous analyses using chiral perturbation theory~\cite{Villadoro:2006nj,Lacour:2007wm,Geng:2014efa,Anderson:1993as,Kaiser:2001yc} have typically found SU(3)-breaking effects of the opposite sign. We reiterate that our analysis differs from these works not only in our use of FRR, to which we attribute the stability of our results under the inclusion of decuplet degrees of freedom, but in additional terms which arise at non-zero values of $Q^2$. We interpret these terms, with the free low-energy constants fit to the lattice simulation results, as having partially absorbed the contributions from higher-order effects which are omitted from the conventional chiral expansion.

By providing a framework for the analysis of systematic effects in lattice simulations of the vector form factor, at the simulation values of $Q^2$, this work constitutes progress towards a more precise determination of this quantity from hyperon semileptonic decays. In turn, such a precise determination of $f_1$ could lead to an improved determination of the CKM matrix element $|V_{us}|$ independent of extractions from kaon and tau decays.

\FloatBarrier

\section*{Acknowledgements}

The numerical configuration generation was performed using the BQCD lattice QCD program~\cite{Nakamura:2010qh} on the IBM BlueGeneQ using DIRAC 2 resources (EPCC, Edinburgh, UK), the BlueGene P and Q at NIC (J\"ulich, Germany) and the Cray XC30 at HLRN (Berlin-Hannover, Germany). The BlueGene codes were optimised using Bagel~\cite{Boyle:2009vp}. The Chroma software library~\cite{Edwards:2004sx} was used in the data analysis. This work was supported by the EU grants 283286 (HadronPhysics3), 227431 (Hadron Physics2) and by the University of Adelaide and the Australian
Research Council through the ARC Centre of Excellence for Particle Physics at the Terascale and grants FL0992247 (AWT), DP140103067 (RDY and JMZ), FT120100821 (RDY) and FT100100005 (JMZ).

\appendix

\section{Tables of coefficients}
\label{sec:CoeffTabs}

In this appendix we tabulate the various chiral coefficients which appear in the expressions of section~\ref{sec:ChiPT}.

\begin{table}
\centering
\begin{tabular}{lc}\hline
$B\,B'$ & $f^\text{SU(3)}_{1(B_1B_2)}$ \\ \hline
$\Lambda p$ & $-\sqrt{\frac{3}{2}}$ \\
$\Sigma ^0p$ & $-\frac{1}{\sqrt{2}}$ \\
$\Sigma ^-n$ & $-1$ \\
$\Xi ^0\Sigma ^+$ & $1$ \\
$\Xi ^-\Lambda $ & $\sqrt{\frac{3}{2}}$ \\
$\Xi ^-\Sigma ^0$ & $\frac{1}{\sqrt{2}}$ \\\hline
$\Delta ^{\text{++}}\Sigma ^{*+}$ & $-\sqrt{3}$ \\
$\Delta ^+\Sigma ^{*0}$ & $-\sqrt{2}$ \\
$\Delta ^0\Sigma ^{*-}$ & $-1$ \\
$\Sigma ^{*0}\Xi ^{*-}$ & $-\sqrt{2}$ \\
$\Sigma ^{*+}\Xi ^{*0}$ & $-2$ \\
$\Xi ^{*0}\Omega ^-$ & $-\sqrt{3}$ \\\hline
\end{tabular}
\caption{Coefficients $f^\text{SU(3)}_{1(B_1B_2)}$ for the tree-level vector current transition between octet or decuplet baryon states $B$ and $B'$.}
\end{table}

\begin{table}
\centering
\begin{tabular}{lc}\hline
$\phi\,\phi''$ & $f^\text{SU(3)}_{1(\phi\phi ')}$ \\ \hline
$\pi ^0K^+$ & $-\frac{1}{\sqrt{2}}$ \\
$\pi ^-K^0$ & $-1$ \\
${K^-\pi ^0}$ & $\frac{1}{\sqrt{2}}$ \\
${K^-\eta }$ & $\sqrt{\frac{3}{2}}$ \\
${\overline{K^0}\pi ^+}$ & $1$ \\
${\eta K^+}$ & $-\sqrt{\frac{3}{2}}$ \\\hline
\end{tabular}
\caption{Coefficients $f^\text{SU(3)}_{1(\phi\phi ')}$ for the tree-level vector current transition between meson states $\phi$ and $\phi'$.}
\end{table}

\begin{table}
\centering
\begin{tabular}{lc}\hline
$B\,B'$ & $C_{BB'}$ \\ \hline
${\text{$\Lambda $p}}$ & $-\frac{{b_D}+3 {b_F}}{\sqrt{6}}$ \\
${\Sigma ^0p}$ & $\frac{{b_D}-{b_F}}{\sqrt{2}}$ \\
${\Sigma ^-n}$ & ${b_D}-{b_F}$ \\
${\Xi ^0\Sigma ^+}$ & ${b_D}+{b_F}$ \\
${\Xi ^-\Lambda }$ & $-\frac{{b_D}-3 {b_F}}{\sqrt{6}}$ \\
${\Xi ^-\Sigma ^0}$ & $\frac{{b_D}+{b_F}}{\sqrt{2}}$ \\ \hline
\end{tabular}
\caption{Coefficients $C_{BB'}$ for the vector current transition between baryon states $B$ and $B'$.}
\end{table}

\begin{table}\centering
\begin{tabular}{lcccccc}\hline
\multicolumn{7}{c}{$C_{B_1B_2\phi V}$}\\
 & $\Lambda p$ & $\Sigma^0 p$ & $\Sigma^- n$ & $\Xi^0 \Sigma^+$ & $\Xi^- \Lambda$ & $\Xi^- \Sigma^0$ \\\hline
$\pi^0$ &$\frac{\sqrt{\frac{3}{2}}}{8}$ & $\frac{1}{8 \sqrt{2}}$ & $\frac{1}{8}$ & $-\frac{1}{8}$ & $-\frac{\sqrt{\frac{3}{2}}}{8}$ & $-\frac{1}{8 \sqrt{2}}$ \\
$\pi^+$ & $\frac{\sqrt{\frac{3}{2}}}{8}$ & $\frac{1}{8 \sqrt{2}}$ & $\frac{1}{8}$ & $-\frac{1}{8}$ & $-\frac{\sqrt{\frac{3}{2}}}{8}$ & $-\frac{1}{8 \sqrt{2}}$ \\
$\pi^-$ & $\frac{\sqrt{\frac{3}{2}}}{8}$ & $\frac{1}{8 \sqrt{2}}$ & $\frac{1}{8}$ & $-\frac{1}{8}$ & $-\frac{\sqrt{\frac{3}{2}}}{8}$ & $-\frac{1}{8 \sqrt{2}}$ \\
$K^0$ & $\frac{\sqrt{\frac{3}{2}}}{8}$ & $\frac{1}{8 \sqrt{2}}$ & $\frac{1}{8}$ & $-\frac{1}{8}$ & $-\frac{\sqrt{\frac{3}{2}}}{8}$ & $-\frac{1}{8 \sqrt{2}}$ \\
$K^+$ & $\frac{\sqrt{\frac{3}{2}}}{4}$ & $\frac{1}{4 \sqrt{2}}$ & $\frac{1}{4}$ & $-\frac{1}{4}$ & $-\frac{\sqrt{\frac{3}{2}}}{4}$ & $-\frac{1}{4 \sqrt{2}}$ \\
$K^-$ & $\frac{\sqrt{\frac{3}{2}}}{4}$ & $\frac{1}{4 \sqrt{2}}$ & $\frac{1}{4}$ & $-\frac{1}{4}$ & $-\frac{\sqrt{\frac{3}{2}}}{4}$ & $-\frac{1}{4 \sqrt{2}}$ \\
$\overline{K}^0$ & $\frac{\sqrt{\frac{3}{2}}}{8}$ & $\frac{1}{8 \sqrt{2}}$ & $\frac{1}{8}$ & $-\frac{1}{8}$ & $-\frac{\sqrt{\frac{3}{2}}}{8}$ & $-\frac{1}{8 \sqrt{2}}$ \\
$\eta$ & $\frac{3 \sqrt{\frac{3}{2}}}{8}$ & $\frac{3}{8 \sqrt{2}}$ & $\frac{3}{8}$ & $-\frac{3}{8}$ & $-\frac{3 \sqrt{\frac{3}{2}}}{8}$ & $-\frac{3}{8 \sqrt{2}}$ \\\hline
\end{tabular}
\caption{Coefficients $C_{BB'\phi V}$ for the tree-level vector current transition between baryon states $B$ and $B'$ with the emission of a meson $\phi$.}
\end{table}

\bibliography{TransitionBib}

\end{document}